\documentclass[12pt]{article}
\usepackage[dvips]{graphicx}
\usepackage{axodraw,epsfig,psfrag,amsmath,amssymb,cite}

\psfigdriver{dvips}
\newcommand{\MNS}{{\text{MNS}}}
\newcommand{\eV}{\text{eV}}
\newcommand{\GeV}{\text{GeV}}
\newcommand{\Hlilj}{H^{\pm\pm}\to l_i^\pm l_j^\pm}

\newcommand{\BR}{\text{BR}}
\newcommand{\BRll}{\text{BR}( H^{\pm\pm} \to l^\pm l^\pm )}
\newcommand{\BRlili}{\text{BR}( H^{\pm\pm} \to l_i^\pm l_i^\pm )}
\newcommand{\BRlilj}{\text{BR}( H^{\pm\pm} \to l_i^\pm l_j^\pm )}

\newcommand{\BRmm}{\text{BR}( H^{\pm\pm} \to \mu^\pm \mu^\pm )}

\newcommand{\rel}{{\text{rel}}}
\newcommand{\true}{{\text{true}}}
\newcommand{\fit}{{\text{fit}}}

\newcommand{\eff}{{\text{eff}}}
\newcommand{\BG}{{\text{BG}}}
\newcommand{\CPC}{{\text{CPC case}}}
\newcommand{\CPV}{{\text{CPV}}}
\newcommand{\allow}{{\text{allow}}}
\newcommand{\hierarchy}{\text{hierarchy}}

\renewcommand{\thefootnote}{\fnsymbol{footnote}}

\newcommand{\prepr}[1] {\begin{flushright}  {\bf #1} \end{flushright} \vskip 1.cm}
\newcommand{\titul}[1] {\boldmath \begin{center}{\Large {\bf #1 } } 
\end{center}
\vskip 0.8cm}
\newcommand{\autor}[1] {\begin{center}  {\bf \lineskip .3cm #1  }
                        \end{center} }
\newcommand{\place}[1] {\begin{center}  {\normalsize \bf \it #1   } \end{center}}
\newcounter{muni}

\def\bmaT{\left(\begin{array}{ccc}}
\def\emaT{\end{array}\right)}
\def\bma{\left( \begin{array} }
\def\ema{\end{array} \right)}
\def\gsim{~{\rlap{\lower 3.5pt\hbox{$\mathchar\sim$}}\raise 1pt\hbox{$>$}}\,}
\def\lsim{~{\rlap{\lower 3.5pt\hbox{$\mathchar\sim$}}\raise 1pt\hbox{$<$}}\,}

\makeatletter
\def\fmslash{\@ifnextchar[{\fmsl@sh}{\fmsl@sh[0mu]}}
\def\fmsl@sh[#1]#2{%
  \mathchoice
    {\@fmsl@sh\displaystyle{#1}{#2}}%
    {\@fmsl@sh\textstyle{#1}{#2}}%
    {\@fmsl@sh\scriptstyle{#1}{#2}}%
    {\@fmsl@sh\scriptscriptstyle{#1}{#2}}}
\def\@fmsl@sh#1#2#3{\m@th\ooalign{$\hfil#1\mkern#2/\hfil$\crcr$#1#3$}}
\makeatother
\begin{document}
\hbadness=10000
\pagenumbering{arabic}
\begin{titlepage}
\prepr{SISSA 99/2007/EP\\
\hspace{30mm} December 2007}
\begin{center}
\titul{\bf Probing Majorana Phases and Neutrino Mass Spectrum\\
in the Higgs Triplet Model at the LHC}

\autor{A.G. Akeroyd$^{1,2}$\footnote{akeroyd@mail.ncku.edu.tw}, 
Mayumi Aoki$^3$\footnote{mayumi@icrr.u-tokyo.ac.jp}
and Hiroaki Sugiyama$^4$\footnote{sugiyama@sissa.it}}
\place{1: Department of Physics,\\
National Cheng Kung University, Tainan, 701 Taiwan}
\place{2: National Center for Theoretical Sciences,\\
Taiwan}
\place{3: ICRR, University of Tokyo, Kashiwa 277-8582, Japan}

\place{4: SISSA, via Beirut 2-4, I-34014 Trieste, Italy}
%\place{4: Theory Group, KEK, 1-1 Oho, \\
%Tsukuba, Ibaraki, 305-0801 Japan}

\end{center}

\vskip2.0cm

\begin{abstract}
\noindent
Doubly charged Higgs bosons ($H^{\pm\pm}$) are a distinctive
signature of the Higgs Triplet Model
of neutrino mass generation. If $H^{\pm\pm}$ is
relatively light ($m_{H^{\pm\pm}}< 400$ GeV) it will be produced
copiously at the LHC, which could enable precise measurements of
the branching ratios of the decay channels  
$H^{\pm\pm}\to l^\pm_i l^\pm_j$. Such branching ratios are
determined solely by the neutrino mass matrix 
which allows the model to be tested at the LHC.
We quantify the dependence of the leptonic branching ratios
on the absolute neutrino mass and Majorana phases, and 
present the permitted values for the channels 
$e^\pm e^\pm,e^\pm\mu^\pm$ and $\mu^\pm\mu^\pm$.
It is shown that precise measurements of
these three branching ratios are sufficient 
to extract information on the neutrino mass spectrum and 
probe the presence of CP violation from Majorana phases.

\end{abstract}

\vskip1.0cm
{\bf  PACS index :12.60.Fr,14.80.Cp,14.60.Pq}
\end{titlepage}
\newpage

\pagestyle{plain}
\renewcommand{\thefootnote}{\arabic{footnote} }
\setcounter{footnote}{0}

\section{Introduction}

The firm evidence from a variety of experiments that neutrinos
oscillate and possess
a small mass below the eV scale necessitates physics beyond the
Standard Model (SM). 
Consequently models of neutrino mass generation which can be
probed at present and forthcoming experiments are of great
phenomenological interest. In particular, those models 
which can provide a distinctive experimental 
signature, such as a New Physics particle
with a mass of the TeV scale or less, are especially appealing
in light of the approaching commencement of the CERN Large 
Hadron Collider (LHC). 

Doubly charged Higgs bosons ($H^{\pm\pm}$) arise in a variety 
of models of neutrino mass generation as 
members of $I=1$, $Y=2$ scalar triplets%
~\cite{Mohapatra:1979ia, Magg:1980ut, Schechter:1980gr, Cheng:1980qt,
Georgi:1985nv, Ma:2000wp, Chun:2003ej, Chen:2005mz, Hung:2006ap,
 Sahu:2007uh, Accomando:2006ga}
and $I=0$, $Y=4$ scalar singlets~\cite{I0Y4}.
Such particles can be relatively light (i.e., with masses 
of the electroweak scale) 
and have impressive discovery potential
at hadron colliders due to their low background signature 
$\Hlilj$ and sizeable
cross-sections. The ongoing searches at the Fermilab Tevatron 
\cite{Abazov:2004au, Acosta:2004uj} anticipate 
sensitivity to $m_{H^{\pm\pm}}< 250$~GeV for the decay channel
$\Hlilj$ ($i, j=e,\mu$) with the expected final integrated
luminosities of up to 8~fb$^{-1}$. 
LHC simulations~\cite{Azuelos:2005uc, Rommerskirchen:2007jv}
show that discovery for $m_{H^{\pm\pm}}< 1$~TeV 
is possible with $300$~fb$^{-1}$,
and as little as 1~${\rm fb}^{-1}$ is needed to 
probe $m_{H^{\pm\pm}}< 400$~GeV\@. 

Discovery of $H^{\pm\pm}$ with $m_{H^{\pm\pm}}< 400$~GeV would enable 
precise measurements of the branching ratios (BRs) of
$\Hlilj$ with the anticipated final
integrated luminosity at the LHC\@. Models which predict  
$\BRlilj$ in terms of experimentally constrained
and/or measured parameters are of particular phenomenological interest.
In the Higgs Triplet Model (HTM)~\cite{Schechter:1980gr},\cite{Cheng:1980qt}
neutrinos acquire a Majorana mass
given by the product of a triplet Yukawa coupling ($h_{ij}$)
and a triplet vacuum expectation value $v_\Delta$. 
Consequently in the HTM there is a direct connection 
between $h_{ij}$ and the neutrino mass matrix which gives rise to 
phenomenological predictions for processes which depend on $h_{ij}$.
Since the coupling $h_{ij}$ determines $\BRlilj$, 
this mechanism of neutrino mass generation can be tested if 
precise measurements of $\BRlilj$ are available%
~\cite{Ma:2000wp, Chun:2003ej}.
A detailed quantitative study of the dependence of $h_{ij}$ on all the 
neutrino oscillation parameters has not yet been performed
(for previous analyses see \cite{Chun:2003ej, Akeroyd:2005gt}).

 Of particular interest is the dependence of
$h_{ij}$ on the absolute neutrino mass and Majorana phases
which is the focus of the present work.
 Those parameters, which cannot be probed in neutrino
oscillation experiments, would significantly affect 
$\BRlilj$.
We perform a study of the
capability of the LHC to probe the
neutrino mass spectrum and Majorana phases 
assuming that neutrino mass is
generated solely by the combination $h_{ij}v_\Delta$ in the HTM\@.
In particular,
we investigate the possibility
to establish $m_0\neq 0$ and/or
CP-violation from Majorana phases at the LHC by means of
a $\chi^2$ analysis with three $2l$ channels of $H^{\pm\pm}$ decays.
It is extremely difficult for neutrinoless double beta decay experiments%
~\cite{Avignone:2007fu}
to measure CP-violation from Majorana phases
because they affect this process in combination
with unmeasured parameters~\cite{phase-0nbb},
while the absolute neutrino mass can only be measured directly
by the future Tritium beta decay experiment~\cite{Bornschein:2003xi}
if $m > 0.2\eV$.

Our work is organized as follows: in section 2 we briefly review 
the HTM; section 3 describes the phenomenology of $H^{\pm\pm}$
at hadron colliders; the numerical analysis is contained in
section 4 with details of a $\chi^2$ analysis presented in 
the appendix;
conclusions are given in section 5.

\section{The Higgs Triplet Model}

In the Higgs Triplet Model~(HTM)~\cite{Schechter:1980gr},\cite{Cheng:1980qt}
a $I=1,Y=2$ complex $SU(2)_L$ isospin triplet of 
scalar fields is added to the SM Lagrangian. 
Such a model can provide a Majorana mass for the observed neutrinos 
without the introduction of a right-handed neutrino via the 
gauge invariant Yukawa interaction:
\begin{equation}
{\cal L}=h_{ij}\psi_{iL}^TCi\tau_2\Delta\psi_{jL}+h.c
\label{trip_yuk}
\end{equation}
Here $h_{ij} (i,j=1,2,3)$ is a complex
and symmetric coupling,
$C$ is the Dirac charge conjugation operator, $\tau_2$
is a Pauli matrix,
$\psi_{iL}=(\nu_i, l_i)_L^T$ is a left-handed lepton doublet,
and $\Delta$ is a $2\times 2$ representation of the $Y=2$
complex triplet fields:
\begin{equation}
\Delta
=\bma{cc}
\Delta^+/\sqrt{2}  & \Delta^{++} \\
\Delta^0       & -\Delta^+/\sqrt{2}
\ema
\end{equation}
A non-zero triplet vacuum expectation value $\langle\Delta^0\rangle$ 
gives rise to the following mass matrix for neutrinos:
\begin{equation}
m_{ij}=2h_{ij}\langle\Delta^0\rangle = \sqrt{2}h_{ij}v_{\Delta}
\label{nu_mass}
\end{equation}
The necessary non-zero $v_{\Delta}$ arises from the minimization of
the most general $SU(2)\otimes U(1)_Y$ invariant Higgs potential,
which is written as follows~\cite{Ma:2000wp, Chun:2003ej}
(with $\Phi=(\phi^+,\phi^0)^T$):
\begin{eqnarray}
V&=&m^2(\Phi^\dagger\Phi)+\lambda_1(\Phi^\dagger\Phi)^2+M^2
{\rm Tr}(\Delta^\dagger\Delta) +
\lambda_2[{\rm Tr}(\Delta^\dagger\Delta)]^2+ \lambda_3{\rm Det}
(\Delta^\dagger\Delta)  \nonumber \\
&&+\lambda_4(\Phi^\dagger\Phi){\rm Tr}(\Delta^\dagger\Delta)
+\lambda_5(\Phi^\dagger\tau_i\Phi){\rm Tr}(\Delta^\dagger\tau_i
\Delta)+\left(
{1\over \sqrt 2}\mu(\Phi^Ti\tau_2\Delta^\dagger\Phi) + h.c \right)
\label{higgs_potential}
\end{eqnarray}
Here $m^2<0$ in order to ensure $\langle\phi^0\rangle=v/\sqrt 2$ which
spontaneously breaks $SU(2)\otimes U(1)_Y$
to  $U(1)_Q$, and $M^2\,(>0)$ is the mass term for the triplet scalars.
In the model of Gelmini-Roncadelli~\cite{Gelmini:1980re} 
the term $\mu(\Phi^Ti\tau_2\Delta^\dagger\Phi)$ is absent,
which leads to spontaneous violation of lepton number for $M^2<0$.
The resulting Higgs
spectrum contains a massless triplet scalar (majoron, $J$) and another light 
scalar ($H^0$). Pair production via $e^+e^-\to H^0J$ would give a large 
contribution to the invisible width of the $Z$ and this model
was excluded at LEP\@. 
The inclusion of the term $\mu(\Phi^Ti\tau_2\Delta^\dagger\Phi$)
explicitly breaks lepton number when $\Delta$ is assigned $L=2$, and eliminates
the majoron~\cite{Schechter:1980gr, Cheng:1980qt}.
 Thus the scalar potential in
eq.~(\ref{higgs_potential}) together with the triplet Yukawa interaction of
eq.~(\ref{trip_yuk}) lead to a phenomenologically viable model of neutrino mass
generation. The expression for $v_\Delta$
resulting from the minimization of $V$ is:
\begin{equation}
v_\Delta \simeq \mu v^2/2M^2
\label{triplet_vev}
\end{equation} 
In the scenario of light triplet scalars ($M\approx v$) 
within the discovery reach of
the LHC, eq.~(\ref{triplet_vev}) leads to $v_\Delta\approx \mu$.
In extensions of the HTM the term $\mu(\Phi^Ti\tau_2\Delta^\dagger\Phi$) 
may arise in various ways: i) the vev of a Higgs singlet field%
~\cite{Schechter:1981cv, Diaz:1998zg}; 
ii) be generated at higher orders in perturbation theory~\cite{Chun:2003ej};
iii) be generated in the effective Lagrangian \cite{Sahu:2007uh};
%with the possibility of $\mu>> v_\Delta$ \cite{Sahu:2007uh}; 
iv) originate in the context of extra dimensions%
~\cite{Ma:2000wp, Chen:2005mz}.

An upper limit on $v_\Delta$ can be obtained from
considering its effect on the parameter $\rho(=M^2_W/M_Z^2\cos^2\theta_W)$. 
In the SM $\rho=1$ at tree-level, while in the HTM one has
(where $x=v_\Delta/v$):
\begin{equation}
\rho\equiv 1+\delta\rho={1+2x^2\over 1+4x^2}
\label{deltarho}
\end{equation}
The measurement $\rho\approx 1$ leads to the bound
$v_\Delta/v\lsim 0.03$, or  $v_\Delta<8$~GeV\@.
At the 1-loop level $v_\Delta$ must be renormalized and explicit
analyses lead to bounds on its magnitude similar to those derived from
the tree-level analysis~\cite{tree}.

The HTM has seven Higgs bosons $(H^{++},H^{--},H^+,H^-,H^0,A^0,h^0)$.
The doubly charged $H^{\pm\pm}$ is entirely composed of the triplet 
scalar field $\Delta^{\pm\pm}$, 
while the remaining eigenstates are in general mixtures of the  
doublet and triplet fields. Such mixing is proportional to the 
triplet vev, and hence small {\it even if} $v_\Delta$
assumes its largest value of a few GeV\@.
Therefore $H^\pm,H^0,A^0$ are predominantly composed 
of the triplet fields, while $h^0$ is predominantly composed of the 
doublet field and plays the role of the SM Higgs boson.
The masses of $H^{\pm\pm},H^\pm,H^0,A^0$ are of order $M$ with splittings 
of order $\lambda_5v$.
For $M< 1$~TeV of interest for direct searches for the Higgs 
bosons at the LHC, the couplings $h_{ij}$ are
constrained to be ${\cal O}(0.1)$ or less by a variety of processes 
such as $\mu\to eee, \tau\to lll$ etc.\ which are reviewed in 
\cite{Gunion:1989in, Cuypers:1996ia}.

\section{Production and Decay of $H^{\pm\pm}$ at Hadron Colliders}

The most distinct and experimentally accessible decay mode
of $H^{\pm\pm}$ is to two same-sign charged leptons~\cite{Rizzo:1981xx}.
Without loss of generality one can work in the basis in which the  
charged lepton mass matrix is diagonal i.e., $l^\pm_i$ are the mass
eigenstates.
Then the decay rate for $\BRlilj$
($i,j=e,\mu,\tau$)
is given by:
\begin{equation}
\Gamma(H^{\pm\pm}\to l_i^\pm l_j^\pm)=
S{m_{H^{\pm\pm}}\over 8\pi}|h_{ij}|^2
\label{Hlldecay}
\end{equation}
where $S=1(2)$ for $i=j$ ($i\ne j$). 
Clearly $\Gamma(H^{\pm\pm}\to l_i^\pm l_j^\pm)$ 
depends crucially on the {\sl absolute} values of the
$h_{ij}$, where $h_{ij}$ is related to the neutrino mass
matrix via eq.~(\ref{nu_mass}). 
However, if no other decay modes for
$H^{\pm\pm}$ are open kinematically the leptonic BRs are determined 
solely by the {\sl relative} values of $h_{ij}$.
Other decay modes for $H^{\pm\pm}$ can be important in regions of 
parameter space e.g., i) $H^{\pm\pm}\to H^\pm W^*$, which
is potentially sizeable for $m_{H^{\pm\pm}} > m_{H^\pm}$%
~\cite{Akeroyd:2005gt, Chakrabarti:1998qy}, and
ii)  $H^{\pm\pm}\to W^\pm W^\pm$, which is proportional
to the triplet vev, $v_{\Delta}$.
 In this work,
we assume $m_{H^{\pm}} \geq  m_{H^{\pm\pm}}$
(which precludes $H^{\pm\pm}\to H^\pm W^*$) 
and $v_\Delta < 1 \,{\rm MeV}$, which suppresses $H^{\pm\pm}\to W^\pm W^\pm$
sufficiently in the HTM (e.g., see \cite{Han:2007bk}).
 Then,
$\Hlilj$ can be regarded as
the sole decay mode for $H^{\pm\pm}$
and one has:
\begin{eqnarray}
\BR_{ll^\prime} \equiv
\BRlilj = \frac{S |h_{ij}|^2}{\sum_{ij} |h_{ij}|^2}.
\label{BRll}
\end{eqnarray}

\subsection{Searches for $H^{\pm\pm}$ at the Tevatron}

In the year 2003 the Fermilab Tevatron performed the 
first search for $H^{\pm\pm}$
at a hadron collider.\footnote{Direct searches for 
$H^{\pm\pm}$ have also been performed at
LEP~\cite{LEP} and HERA~\cite{Aktas:2006nu}.}
The D0 collaboration~\cite{Abazov:2004au} 
searched for $H^{\pm\pm}\to \mu^\pm\mu^\pm$ 
while the CDF collaboration \cite{Acosta:2004uj}
searched for 3 final states: $H^{\pm\pm}\to e^\pm e^\pm,
e^\pm \mu^\pm, \mu^\pm\mu^\pm$. The assumed production mechanism for
$H^{\pm\pm}$ is $q\overline q\to \gamma^*,Z^*\to H^{++}H^{--}$, which
proceeds via gauge strength couplings and 
depends on only one unknown parameter, $m_{H^{\pm\pm}}$%
~\cite{Gunion:1989in, Huitu:1996su}.
\begin{center}
\begin{picture}(220,90)(-10,-20) % y_2 controls equation position
\Photon(60,25)(118,25){4}{8}
\ArrowLine(60,25)(10,55)
\ArrowLine(10,-5)(60,25)
\DashLine(168,-5)(118,25){3}
\DashLine(118,25)(168,55){3}
\Text(2,55)[]{$\overline q$}
\Text(2,-5)[]{$q$}
\Text(189,-2)[]{$H^{--}$}
\Text(189,60)[]{$H^{++}$}
\Text(90,38)[]{$\gamma,Z$}
\Vertex(118,25){3}
\end{picture}
\hspace*{1cm} 
\begin{picture}(220,90)(-10,-20) % y_2 controls equation position
\Photon(60,25)(118,25){4}{8}
\ArrowLine(60,25)(10,55)
\ArrowLine(10,-5)(60,25)
\DashLine(168,-5)(118,25){3}
\DashLine(118,25)(168,55){3}
\Text(2,55)[]{$\overline q$}
\Text(2,-5)[]{$q'$}
\Text(189,-2)[]{$H^{\mp}$}
\Text(189,60)[]{$H^{\pm\pm}$}
\Text(90,38)[]{$W^\pm$}
\Vertex(118,25){3}
\end{picture}

\end{center}
The searches performed in \cite{Abazov:2004au, Acosta:2004uj} 
seek at least
one pair of same-sign leptons with high invariant mass i.e.,
the search is sensitive to {\it single production}
of $H^{\pm\pm}$. The SM background
can be reduced to negligible proportions with suitable cuts. 
Single $H^{\pm\pm}$ production
mechanisms which involve a dependence on potentially small parameters such as
the Yukawa coupling $h_{ij}$ or triplet vev $v_{\Delta}$  
are subdominant at Tevatron energies (e.g., see \cite{Maalampi:2002vx}).
In \cite{Akeroyd:2005gt}
it was suggested that this search strategy is also sensitive to
the mechanism $q'\overline q\to W^*\to H^{\pm\pm}H^\mp$%
~\cite{Dion:1998pw}, which has
a cross-section comparable in magnitude to that of  
$q\overline q\to \gamma^*,Z^*\to H^{++}H^{--}$.
The following inclusive single $H^{\pm\pm}$ cross-section was introduced, 
which would extend the search sensitivity to larger values of
$m_{H^{\pm\pm}}$ and strengthen the mass limits on $m_{H^{\pm\pm}}$ 
derived in \cite{Abazov:2004au, Acosta:2004uj}:
\begin{equation}
\sigma_{H^{\pm\pm}}= \sigma(q\overline q\to \gamma^*,Z^*\to H^{++}H^{--})
+2\sigma(q\overline q\to W^*\to H^{++}H^-)
\label{single_prod}
\end{equation} 
Here the factor of 2 accounts for the CP conjugate process
$q\overline q\to W^*\to H^{--}H^+$.

In 2006 the CDF collaboration searched for
$H^{\pm\pm}$ decays involving $\tau^\pm$~\cite{Safonov:2006hg}.
The strategy of searching for one pair of same-sign leptons ($2l$) is not  
effective due to the larger SM backgrounds,
and instead three ($3l$) and four ($4l$) lepton searches were performed.
The production mechanism $q\overline q\to \gamma^*,Z^*\to H^{++}H^{--}$
was assumed. The process $q\overline q\to W^*\to H^{\pm\pm}H^\mp$
never contributes to the $4l$ signature, but can contribute to the
$3l$ signature if $H^\pm$ decays leptonically, $H^\pm\to l^\pm\nu$. 

In Table~\ref{mass_limits} the mass limits for $m_{H^{\pm\pm}}$ from the
Tevatron searches are summarized. A blank entry signifies that
no search has yet been performed. The displayed mass limits assume 
production via $q\overline q\to \gamma^*,Z^*\to H^{++}H^{--}$ for
a $H^{\pm\pm}$ belonging to a $SU(2)_L$ triplet with $Y=2$.
Moreover, $\BRlilj=100\%$ in a given channel
is assumed.
The ultimate sensitivity at the Tevatron is expected to be
$m_{H^{\pm\pm}}\sim 250$~GeV in the $ee,e\mu$ and $\mu\mu$ channels.
\begin{table}
\begin{center}
\begin{tabular}{|c|c|c|c|c|c|c|}
\hline
  & $ee$ & $e\mu$ & $\mu\mu$ & $e\tau$ & $\mu\tau$ & $\tau\tau$ \tabularnewline
\hline
2l&  $>133$ GeV & $> 113$ GeV  & $>136$ GeV  &   &  & 
\tabularnewline
\hline
3l&  &  &  & $>114$ GeV  & $>112$ GeV & \tabularnewline
\hline
4l& &  &  &  $>114$ GeV  &  $>112$ GeV &\tabularnewline
\hline
\end{tabular}
\end{center}
\caption{Mass limits on $m_{H^{\pm\pm}}$ from searches
for $\Hlilj$ at Tevatron Run II.}
\label{mass_limits}
\end{table}

\subsection{Simulations of $H^{\pm\pm}$ production at the LHC}

Several simulations have been performed for $\Hlilj$ 
($i, j=e,\mu,\tau$)
at the LHC~\cite{Azuelos:2005uc, Rommerskirchen:2007jv, Han:2007bk,
LHCH++, Allanach:2006fy}.
The production mechanism is assumed to 
be $q\overline q\to \gamma^*,Z^*\to H^{++}H^{--}$ followed by
$H^{++}H^{--}\to llll$. The LHC sensitivity to 
$\Hlilj$ 
considerably extends that at the Tevatron
due to the increased cross-sections and larger luminosities
e.g., the analysis of \cite{Rommerskirchen:2007jv} shows that  
a $H^{\pm\pm}$ can be discovered for $m_{H^{\pm\pm}}< 800$~GeV assuming 
$\BRmm =100\%$ and ${\cal L}=50$~fb$^{-1}$.
Importantly, all the above simulations suggest that 
as little as 1~fb$^{-1}$ is needed for discovery of
$m_{H^{\pm\pm}}<400$~GeV
if one of
BR($H^{\pm\pm}\to e^\pm e^\pm, e^\pm \mu^\pm, \mu^\pm \mu^\pm$)
is large, and 
hence such a light $H^{\pm\pm}$ would be found very quickly at the LHC\@.

The sensitivity of the LHC to single production of $\Hlilj$ for
$i, j=e,\mu$ has only been performed in \cite{Allanach:2006fy}, and importantly 
the SM background was shown to be negligible in the signal region of
high invariant mass.
It was concluded that such a search strategy allows more $\Hlilj$ 
events than the 4 lepton search since the event number is linear 
(and not quadratic) in $\BRlilj$. Therefore the 
$2l$ search is more effective
at probing small $\BRlilj$. 
Importantly, the addition of the channel
$q\overline q\to W^*\to H^{\pm\pm}H^\mp$ (eq.~(\ref{single_prod}))
would further enhance the event number for a given $m_{H^{\pm\pm}}$.

In Table~\ref{event_no} we show approximate expected 
numbers of $2l$ and $4l$ events arising from pair and singly produced
$\Hlilj$ at the LHC\@. We only consider the decay channels
$H^{\pm\pm}\to e^\pm e^\pm, e^\pm \mu^\pm, \mu^\pm \mu^\pm$
which offer the greatest $H^{\pm\pm}$
discovery potential. A detection efficiency of 0.5 is 
assumed, which is slightly less than the values given in 
\cite{Rommerskirchen:2007jv} for $H^{\pm\pm}\to \mu^\pm \mu^\pm$. 
The theoretical $H^{\pm\pm}$ cross-section is multiplied by 
this detection efficiency and the SM background is taken to be 
negligible. The number of $4l$ events
for a specific $m_{H^{\pm\pm}}$ is denoted by 
$N_{4l}$, assuming integrated luminosities 
of ${\cal L}=30$~fb$^{-1}$ and ${\cal L}=300$~fb$^{-1}$.
The displayed numbers are for $\BRlilj =100\%$
in a given channel. In the HTM, $\BRlilj$
is necessarily $<100\%$ (eq.~(\ref{nu_mass}) and eq.~(\ref{Hlldecay}))
and hence $N_{4l}$ must be multiplied by $[\BRlilj]^2$.
The final
column shows the number of $2l$ events ($N_{2l}$) obtained 
by adding the contribution from the mechanism 
$q\overline q\to W^*\to H^{\pm\pm}H^\mp$
as defined in eq.~(\ref{single_prod}).
We take $m_{H^{\pm\pm}}=m_{H^\pm}$ which increases the 
number of singly produced $H^{\pm\pm}$ events by a factor of around 2.8
for $200~\GeV < m_{H^{\pm\pm}} < 400~\GeV$~\cite{Akeroyd:2005gt}.
For $\BRlilj<100\%$, the numbers presented in Table~\ref{event_no}
are scaled as shown in Appendix~\ref{sec:chi}.

It is clear from Table~\ref{event_no} 
that early discovery of $H^{\pm\pm}$ at the
LHC with $m_{H^{\pm\pm}}<400$~GeV
would allow large event numbers for $H^{\pm\pm}$ with the expected
integrated luminosities of ${\cal L}=300$~fb$^{-1}$.
This would enable precise measurements of 
BR($H^{\pm\pm}\to e^\pm e^\pm, e^\pm \mu^\pm, \mu^\pm \mu^\pm$)
for the dominant channels.
Sensitivity to
BR$(H^{\pm\pm}\to e^\pm e^\pm, e^\pm\mu^\pm, \mu^\pm \mu^\pm)\sim 1\%$
or less would also be possible in the $2l$ channel.

\begin{table}
\begin{center}
\begin{tabular}{|c|c|c|c|}
\hline
$m_{H^{\pm\pm}}$ (GeV) &  $N_{4l}$ (30 fb$^{-1}$) &  $N_{4l}$ 
(300 fb$^{-1}$) & $N_{2l}$ (300 fb$^{-1}$)  
\tabularnewline
\hline
200&  1500 & 15000 &  42000    
\tabularnewline
\hline
300&  300 &  3000 & 8400 \tabularnewline
\hline
400& 90 &  900 & 2500 \tabularnewline
\hline
\end{tabular}
\end{center}
\caption{Approximate number of events for pair production of 
$H^{\pm\pm}$ ($N_{4l}$) and single production 
of $H^{\pm\pm}$ ($N_{2l}$) at the LHC
for integrated luminosities of 
${\cal L}=30$~fb$^{-1}$ and ${\cal L}=300$~fb$^{-1}$
with efficiency $\epsilon_\eff = 0.5$.
We assumed $m_{H^{\pm\pm}} = m_{H^\pm}$
to calculate $N_{2l}$.}
\label{event_no} 
\end{table}

\section{Numerical Analysis}

 The mass matrix for three Dirac neutrinos
is diagonalized by the MNS (Maki-Nakagawa-Sakata) matrix
$V_\MNS$~\cite{Maki:1962mu}
for which the standard parametrization is:
\begin{equation}
V_\MNS^{} =
\bmaT
c_{12}c_{13}                        & s_{12}c_{13}                  & s_{13}e^{-i\delta} \\
-s_{12}c_{23}-c_{12}s_{23}s_{13}e^{i\delta}  & c_{12}c_{23}-s_{12}s_{23}s_{13}e^{i\delta}  & s_{23}c_{13} \\
s_{12}s_{23}-c_{12}c_{23}s_{13}e^{i\delta}   & -c_{12}s_{23}-s_{12}c_{23}s_{13}e^{i\delta} & c_{23}c_{13}  
\emaT
\,,
\end{equation} 
where $s_{ij}\equiv\sin\theta_{ij}$ and $c_{ij}\equiv \cos\theta_{ij}$,
and $\delta$ is the Dirac phase.
 For Majorana neutrinos,
two additional phases appear and then the mixing matrix $V$ becomes
\begin{eqnarray}
 V = V_\MNS \times
     \text{diag}( 1, e^{i\varphi_1 /2}, e^{i\varphi_2 /2}),
\end{eqnarray}
where $\varphi_1$ and $\varphi_2$ are referred to as
the Majorana phases~\cite{Schechter:1980gr, Mphase}.
Since we are working in the basis in which the charged lepton
mass matrix is diagonal, the neutrino mass matrix is then
diagonalized by $V$.
Using eq.~(\ref{nu_mass}) one can write the couplings
$h_{ij}$ as follows~\cite{Ma:2000wp, Chun:2003ej}:
\begin{equation}
h_{ij}=\frac{1}{\sqrt{2}v_\Delta}
\left[
 V_\MNS
 \text{diag}(m_1,m_2 e^{i\varphi_1},m_3 e^{i\varphi_2})
 V_\MNS^T
\right]_{ij}
\label{hij}
\end{equation}
 Then,
eq.~(\ref{BRll}) becomes
\begin{eqnarray}
\BRlilj
= \frac{
   S
   \left\{
   \left[
    V_\MNS
    \text{diag}(m_1,m_2 e^{i\varphi_1},m_3 e^{i\varphi_2})
    V_\MNS^T
   \right]_{ij}
   \right\}^2
  }{
    \sum_k m_k^2
   }.
\end{eqnarray}
Note that the branching ratios
are independent of $m_{H^{\pm\pm}}$ and $v_\Delta$,
and given by neutrino parameters only.

Neutrino oscillation experiments involving solar~\cite{solar}, 
atmospheric~\cite{atm}, accelerator~\cite{acc},
and reactor neutrinos~\cite{reactor}
are sensitive to the mass-squared 
differences and the mixing angles, 
and give the following preferred values:
\begin{eqnarray}
\Delta m^2_{21} \equiv m^2_2 -m^2_1
\simeq 7.9\times 10^{-5} {\rm eV}^2 \,,~~
|\Delta m^2_{31}|\equiv |m^2_3 -m^2_1|
\simeq 2.7\times 10^{-3} {\rm eV}^2\,, \\
\sin^22\theta_{12}\simeq 0.86 \,,~~~~ \sin^22\theta_{23}\simeq 1 \,,~~~~
\sin^22\theta_{13}\lsim 0.13\,.~~~~~~~~~~~~
\label{obs_para}
\end{eqnarray}
The small mixing angle $\theta_{13}$ has not been measured yet
and hence the value of $\delta$ in completely unknown.
 Since the sign of $\Delta m_{31}^2$ is also undetermined at present, 
distinct neutrino mass hierarchy patterns are possible.
The case with $\Delta m^2_{31} >0$ is referred to as
{\it Normal hierarchy} (NH) where $m_1 < m_2 < m_3$
and the case with $\Delta m^2_{31} <0$ is known as
{\it Inverted hierarchy} (IH) where $m_3 <  m_1 < m_2$.  
 Information on the mass $m_0$ of the lightest neutrino
and the Majorana phases
cannot be obtained from neutrino oscillation experiments.
This is because the oscillation probabilities are independent 
of these parameters, not only in vacuum but also in matter.
 If $m_0 \gtrsim 0.2\eV$,
future ${}^3$H beta decay experiment~\cite{Bornschein:2003xi}
can measure it.
Experiments which seek neutrinoless double beta decay%
~\cite{Avignone:2007fu}
are sensitive to only a combination of neutrino masses and phases.
 Certainly, extracting information on Majorana phases alone
from these experiments seems extremely difficult, if not impossible%
~\cite{phase-0nbb}.
Therefore it is worthwhile to consider other possibilities.

Multiplying out eq.~(\ref{hij}) one obtains the following
explicit expressions for $h_{ij}$:
\begin{eqnarray}
h_{ee} &=& \frac{1}{\sqrt{2} v_\Delta}
 \Bigl(
  m_1 c_{12}^2 c_{13}^2
  + m_2 s_{12}^2 c_{13}^2 e^{i\varphi_1}
  + m_3 s_{13}^2 e^{-2i\delta} e^{i\varphi_2}
 \Bigr)\,,
 \nonumber \\
h_{e\mu} &=& \frac{1}{\sqrt{2} v_\Delta}
 \Bigl\{
  m_1 ( -s_{12}c_{23} - c_{12}s_{23}s_{13} e^{i\delta} ) c_{12}c_{13}
 \nonumber \\
 &&\hspace{30mm}
 {}+ m_2 ( c_{12}c_{23} - s_{12}s_{23}s_{13} e^{i\delta} ) s_{12}c_{13} e^{i\varphi_1}
  + m_3 s_{23}c_{13}s_{13} e^{-i\delta} e^{i\varphi_2}
 \Bigr\}\,,
 \nonumber \\
h_{e\tau} &=& \frac{1}{\sqrt{2} v_\Delta}
 \Bigl\{
  m_1 ( s_{12}s_{23} - c_{12}c_{23}s_{13} e^{i\delta} ) c_{12}c_{13}
 \nonumber \\
 &&\hspace{30mm}
  {}+ m_2 ( -c_{12}s_{23} - s_{12}c_{23}s_{13} e^{i\delta} ) s_{12}c_{13} e^{i\varphi_1}
  + m_3 c_{23}c_{13}s_{13} e^{-i\delta} e^{i\varphi_2}
 \Bigr\}\,,
  \nonumber \\
h_{\mu\mu} &=& \frac{1}{\sqrt{2} v_\Delta}
 \Bigl\{
  m_1 ( -s_{12}c_{23} - c_{12}s_{23}s_{13} e^{i\delta} )^2
  + m_2 ( c_{12}c_{23} - s_{12}s_{23}s_{13} e^{i\delta} )^2 e^{i\varphi_1}
  + m_3 s_{23}^2 c_{13}^2 e^{i\varphi_2}
 \Bigr\}\,,
  \nonumber \\
h_{\mu\tau} &=& \frac{1}{\sqrt{2} v_\Delta}
 \Bigl\{
  m_1 ( -s_{12}c_{23} - c_{12}s_{23}s_{13} e^{i\delta} )
        ( s_{12}s_{23} - c_{12}c_{23}s_{13} e^{i\delta} )
 \nonumber \\
 &&\hspace{20mm}
  {}+ m_2 ( c_{12}c_{23} - s_{12}s_{23}s_{13} e^{i\delta} )
        ( -c_{12}s_{23} - s_{12}c_{23}s_{13} e^{i\delta} ) e^{i\varphi_1}
  + m_3 c_{23}s_{23} c_{13}^2 e^{i\varphi_2}
 \Bigr\}\,,
  \nonumber \\
h_{\tau\tau} &=& \frac{1}{\sqrt{2} v_\Delta}
 \Bigl\{
  m_1 ( s_{12}s_{23} - c_{12}c_{23}s_{13} e^{i\delta} )^2
  + m_2 ( -c_{12}s_{23} - s_{12}c_{23}s_{13} e^{i\delta} )^2 e^{i\varphi_1}
  + m_3 c_{23}^2 c_{13}^2 e^{i\varphi_2}
 \Bigr\}\,.
 \nonumber \\
\label{hij_expressions}
\end{eqnarray}
One may express $m_1,m_2,m_3$ in terms of 
two neutrino mass-squared differences ($\Delta m^2_{21},\Delta m^2_{31}$)
and the mass of the lightest neutrino $m_0$.
The $h_{ij}$ are functions of nine parameters:\\
$\Delta m^2_{21}$,$\Delta m^2_{31}$,$m_0$, 
three mixing angles $(\theta_{12},\theta_{13},\theta_{23})$ 
and three complex phases $(\delta, \varphi_1, \varphi_2)$. 
Four of the above  parameters have been measured well 
experimentally (see eq.~\ref{obs_para})
and thus the HTM already provides numerical predictions for 
$\BRlilj$ as a function
of the unmeasured five parameters.
 Four cases
corresponding to no CP violation from Majorana phases 
can be defined as follows: 
Case~I $(\varphi_1=0,\varphi_2=0)$;
Case~II $(\varphi_1=0,\varphi_2=\pi)$;
Case~III $(\varphi_1=\pi,\varphi_2=0)$; 
Case~IV $(\varphi_1=\pi,\varphi_2=\pi)$.
These four cases have been studied
in \cite{Chun:2003ej, Akeroyd:2005gt} for values
of $m_0=0$ or {$\cal O$}(1)~eV\@. In this work we will quantify the 
dependence of $h_{ij}$ on $m_0$, $ \varphi_1$ and $ \varphi_2$.
Those 3 parameters essentially determine $\BRlilj$, with subdominant
corrections from the neutrino oscillation parameters. 
Hence multiple signals for $\Hlilj$
at the LHC would probe $m_0$, $ \varphi_1$ and $ \varphi_2$,
in the context of the HTM\@.

We note here that such collider probes of the Majorana phases
are particular to models in which lepton number violation
(which leads to the Majorana neutrino mass)
is associated with New Physics particles at the TeV scale.
Analogous collider probes are not possible 
in models where the scale of lepton number violation
is much higher e.g., supersymmetric 
models with very heavy right-handed neutrinos with masses of 
order $10^{12}~\GeV$. However, in such models the Majorana phases (which are
also required for successful leptogenesis) 
can significantly affect the rates for the lepton flavour violating (LFV)
decays $\tau\to l\gamma$ and $\mu \to e\gamma$~\cite{LFV1}.
Likewise, in the HTM $ \varphi_1$ and $ \varphi_2$ would
also affect the rates for LFV decays which depend 
explicitly on $h_{ij}$ e.g., $\tau\to lll$ and $\mu \to e\gamma$%
~\cite{Chun:2003ej}. Similar studies of the effects of Majorana phases
on these LFV decays in other models which contain a TeV scale 
$H^{\pm\pm}$ and an analogous $h_{ij}$ coupling have been
performed in \cite{LFV2}.

\begin{figure}[t]
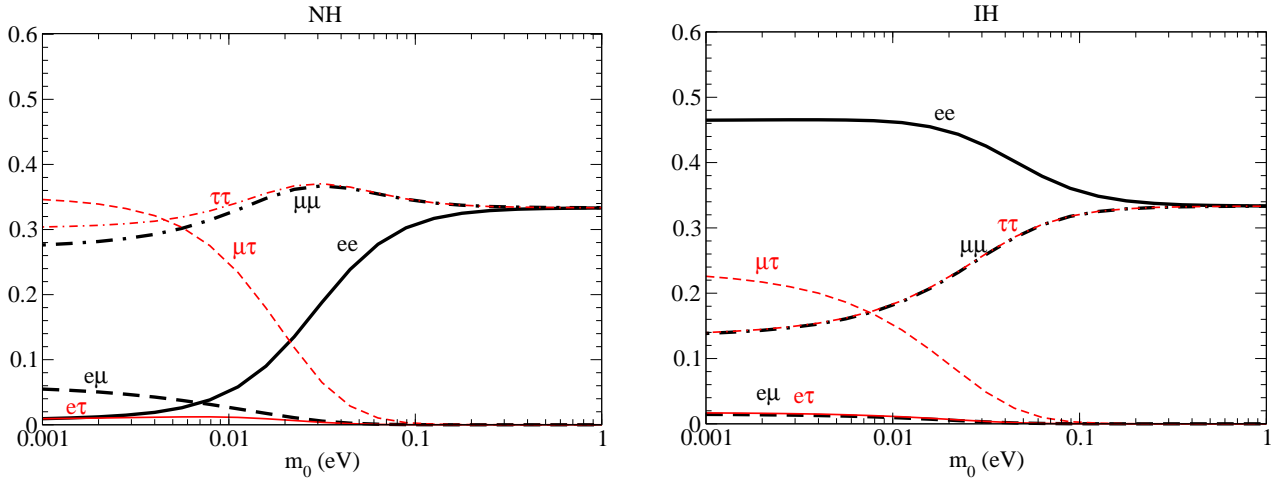

\begin{center}
\includegraphics[width=8cm]{NH.eps} \ \ \ \ \
\includegraphics[width=8cm]{IH.eps}
\caption{%
$\BRlilj$ with no CP violation from Majorana phases
($\varphi_1=\varphi_2=0$) as a function of $m_0$ in 
NH (left) and IH (right).}
\label{fig1}
\end{center}
\end{figure}

\subsection{Dependence of $\BRll$ on the neutrino mass spectrum}

 In Fig.~\ref{fig1},
we plot $\BRlilj$ as a function of $m_0$
for $ \varphi_1=0$ and $ \varphi_2=0$ (Case~I)
in NH and IH\@.
 The values of the oscillation parameters are
fixed in the figures as follows:
\begin{eqnarray}
&&
\Delta m^2_{21}=7.9\times 10^{-5} {\rm eV}^2 \,,~~
|\Delta m^2_{31}|=2.7\times 10^{-3} {\rm eV}^2\,,~~
\sin^22\theta_{12}=0.86\,,
\label{fig1_para1}\\
&&
\sin^22\theta_{23}= 1 \,,~~~~
\sin^22\theta_{13}= 0.13\,,~~~~~ \delta=0.
\label{fig1_para2}
\end{eqnarray}

For NH with smaller $m_0$, the
$e$-related modes are suppressed
because $m_1 (= m_0)$ and $m_2$ are small
and the contribution from the heaviest mass $m_3$
is also small due to the tiny $\theta_{13}$. 
 $\BR_{\mu\mu}$, $\BR_{\mu\tau}$, and ${\BR_{\tau\tau}}$
are roughly equal%
~\footnote{
The na\"{\i}ve expectation is that
$\BR_{\mu\tau}$ is twice as large as the other two modes
because of $|h_{\mu\tau}|^2+|h_{\tau\mu}|^2$.
However, this difference is  accidentally compensated
by the effect of $\Delta m^2_{21}$.
},
and they dominate for smaller $m_0$
because $m_3$ does not appear with $\theta_{13}$.
 Note that
$\BR_{\mu\mu} \simeq \BR_{\tau\tau}$
and $\BR_{e\mu} \simeq \BR_{e\tau}$
can be understood as the approximate symmetry
for $\mu$-$\tau$ exchange by virtue of
$\sin^2{2\theta_{23}}=1$ and tiny $\theta_{13}$.
 For larger $m_0$,
$\BRlili$ dominate
because all coefficients of $m_i$ are positive
for $h_{ii}$ with $\theta_{13} = 0$ in Case~I
and there is no strong cancellation among these terms
even for $\theta_{13}\neq 0$.
 For $m_0 \gtrsim 0.3~\eV$, the
results for NH and IH are almost identical
because of almost degenerate masses.
 Since $\BR$ are given by ratios of $|h_{ij}|^2$,
they converge for large $m_0$.
 This is an attractive feature
because HTM can predict certain ranges of $\BR$
without restricting $m_0$.

On the other hand,
IH case gives rather simple results.
This is because $m_1$ and $m_2$ include the
larger scale $\sqrt{|\Delta m^2_{31}|}$
without any suppression from $\theta_{13}$
and consequently the effects of small $\Delta m^2_{21}$ and $\theta_{13}$
are hidden. In this case,
$\BR_{ee}$ becomes the dominant channel
even for smaller $m_0$;
one na\"{\i}vely expects $\BR_{ee} \simeq 4 \BR_{\mu\mu}$
for $m_0 = 0$ with eq.~(\ref{hij_expressions}).

\begin{figure}[t]
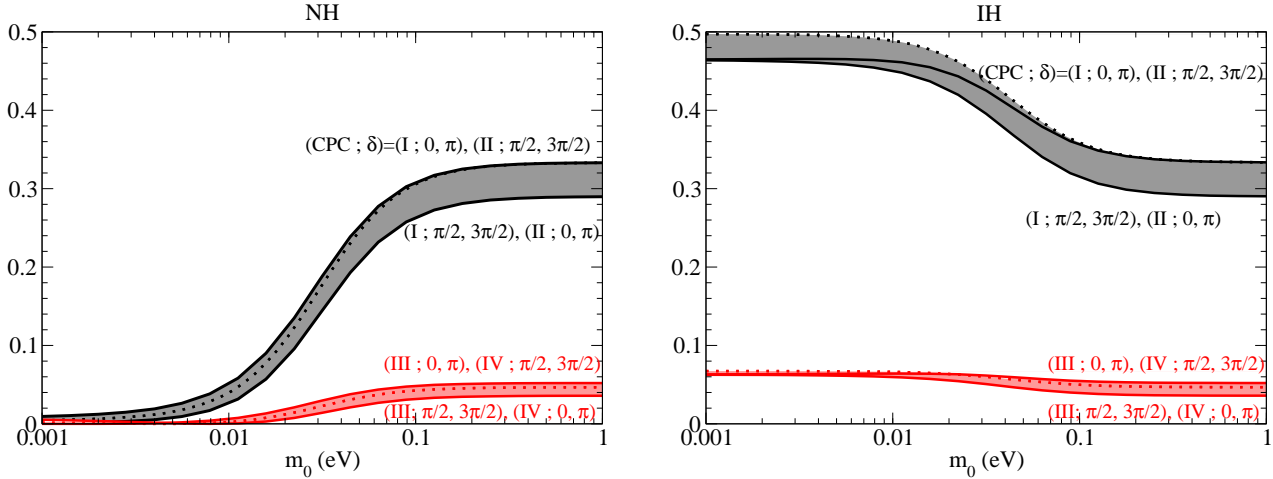

\begin{center}
\includegraphics[width=8cm]{fill_BRee.eps} \ \ \ \ \
\includegraphics[width=8cm]{fill_BRee_IH.eps}
\caption{%
BR$(H^{\pm\pm}\to e^\pm e^\pm)$ as a function of $m_0$ in 
NH (left) and IH (right). 
We take four cases of CP conservation 
(I-IV) for $\sin^22\theta_{13}=0.13$
and 4 values of $\delta$ ($= 0, \pi/2, \pi, 3\pi/2$).
The dotted lines are obtained for $\theta_{13}=0$.}
\label{fig2}
\end{center}
\end{figure}

 In order to quantify the effect of the unmeasured
$(V_\MNS)_{e3}$ on $\BR_{ll^\prime}$,
we show $\BR_{ee}$ for
$\sin^22\theta_{13}=0.13, 0$, and 4 values of 
the Dirac phase $\delta$
($= 0, \pi/2, \pi, 3\pi/2$) for Cases~I to IV
in Fig.~\ref{fig2}.
 Other parameters are same as Fig.~\ref{fig1}.
Several lines are coincident in the figure, 
since $\delta$ and $\varphi_2$ always appear as a combination
$2\delta - \varphi_2$ in $h_{ee}$.
 For example,
Case~I with $\delta = 0$ and Case~II with $\delta = \pi/2$
give the same lines.
Although the contribution of $(V_\MNS)_{e3}$ to $\BR_{ee}$
is considerably smaller than the effect of
varying $m_0$ or the Majorana phases
(as expected by quadratic suppression with small $\theta_{13}$),
it is not negligible.
Consequently, it is also not negligible for other the $\BR$.
 Fig.~\ref{fig2} also shows that the 
HTM predicts small $\BR_{ee}$ in case~III and IV
(cases with $\varphi_1 = \pi$)
for any value of $m_0$, $\theta_{13}$, and $\delta$
in both of NH and IH\@.
 Thus,
Fig.~\ref{fig2} indicates that some information
on Majorana phases may be extracted
without knowledge of the sign of $\Delta m^2_{31}$
and the values of $m_0$, $\theta_{13}$, and $\delta$.

\subsection{Dependence of $\BRlilj$ on Majorana phases}

\begin{figure}[t]
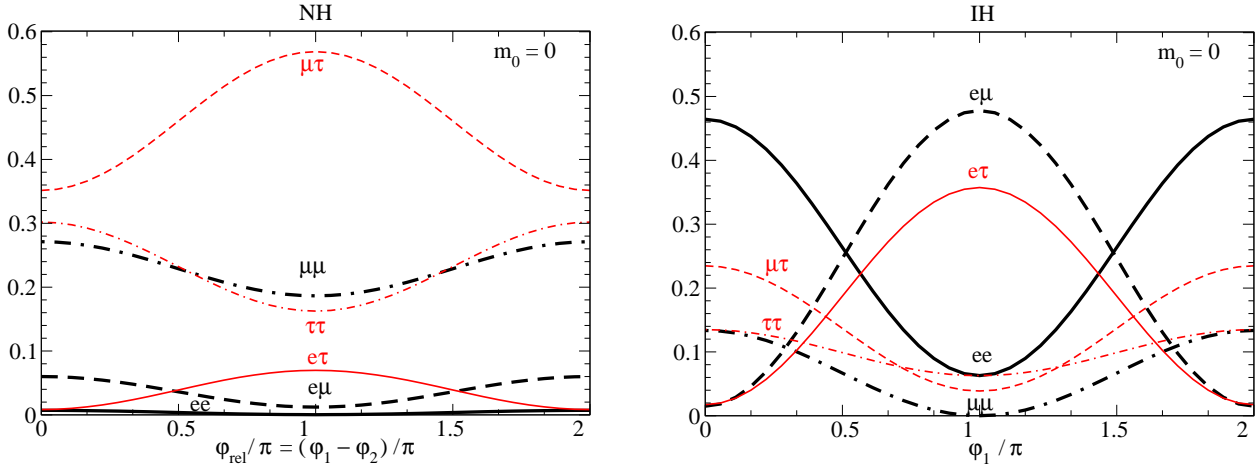

\begin{center}
\includegraphics[width=8cm]{m00n_phirel.eps} \ \ \ \ \
\includegraphics[width=8cm]{m00i_phi1_0.eps}
\caption{%
$\BR_{ll^\prime}$ as a function of $\varphi_{rel}$ in NH (left) and of
$\varphi_1$ in IH (right).}
\label{fig3}
\end{center}
\end{figure}
\begin{figure}[t]
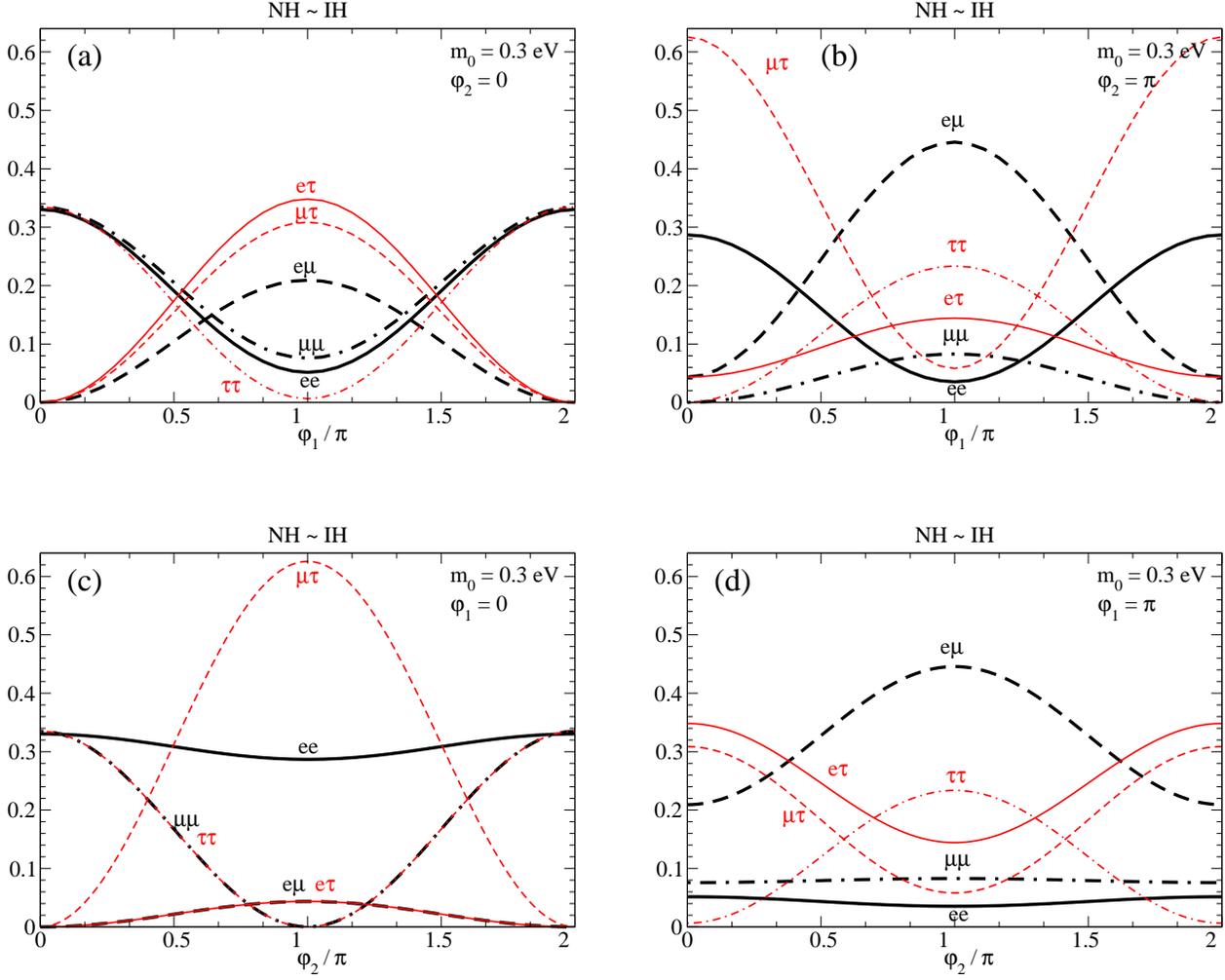

\begin{center}
\includegraphics[width=8cm]{phi1_0.eps} \ \ \ \ \
\includegraphics[width=8cm]{phi1_pi.eps}\\[10mm]
\includegraphics[width=8cm]{phi2_0.eps} \ \ \ \ \
\includegraphics[width=8cm]{phi2_pi.eps}
\caption{%
$\BR_{ll^\prime}$ as a function of 
$\varphi_1$ ($\varphi_2$) with $\varphi_2 = 0$ (a), $\pi$ (b)
($\varphi_1=0$ (c), $\pi$ (d)) for $m_0=0.3\eV$ (NH$\simeq$ IH).}
\label{fig4}
\end{center}
\end{figure}

In this section we show the dependence of $\BRlilj$ on Majorana phases
with the values of the oscillation parameters
given in (\ref{fig1_para1}) and (\ref{fig1_para2}).
 For $m_0=0$,
only the relative phase $\varphi_\rel(= \varphi_1- \varphi_2)$
determines $h_{ij}$ in NH\@.
 In Fig.~\ref{fig3} (left)
we plot $\BRlilj$ as a function of $\varphi_\rel$ for NH with $m_0=0$. 
 Tiny $\theta_{13}$ suppresses the dependence
of $\BR_{ee}$ on $\varphi_\rel$,
and then the HTM gives a clear prediction
of very small $\BR_{ee}$ for this case.
The other $\BR_{ll^\prime}$ change non--negligibly
e.g.,\ $\BR_{e\mu}$ and $\BR_{\mu\mu}$ vary 
by $0.05$ or more, which could be larger than the experimental
error if sufficiently large numbers of $H^{\pm\pm}$ are produced.
In IH for $m_0 = 0$,
$\BR_{ll^\prime}$ does not depend on $\varphi_2$
because it always appears multiplied by $m_0$. 
Fig.~\ref{fig3} (right) shows the dependence of $\BR_{ll^\prime}$
on $\varphi_1$ for this case.
One can see that the dependence on $ \varphi_1$ is even more pronounced
because $m_2$ for $m_0 = 0$ in IH ($= \sqrt{|\Delta m^2_{31}|}$)
is larger than that in NH ($= \sqrt{\Delta m^2_{21}}$).
In particular,
$\BR_{ee}$ and $\BR_{e\mu}$ seem to be very useful
for extracting information on $\varphi_1$
because they have large and opposite dependence.
This dependence on $\varphi_1$
can be understood by the relative sign of
the terms of $m_1$ and $m_2$ in eq.~(\ref{hij_expressions}),
which is $+$ for $h_{ee}$ and $-$ for $h_{e\mu}$,
neglecting $\theta_{13}$.

For $m_0=0.3$~eV,
where neutrino masses are almost degenerate,
% (approximately degenerate neutrinos)
Fig.~\ref{fig4} shows that
the phase $ \varphi_1$ has a large effect on all $\BRlilj$.
 The figure also shows that the
$\varphi_2$ dependence for $\BR_{e\mu}$ and $\BR_{e\tau}$
is sizeable and overcomes the suppression from $\theta_{13}$.
However, the latter suppression factor 
ensures that $\BR_{ee}$ has a very small dependence on $\varphi_2$,
and so this channel is crucial for extracting
clear information on $\varphi_1$ and/or $m_0$
without contamination from the effect of $\varphi_2$.

 It is clear that the Majorana phases have a large effect on
$\BR_{ll^\prime}$ in the HTM, and their inclusion is required
in order to quantify the allowed regions of $\BR_{ll^\prime}$.

\subsection{Sensitivity to $\text{sign}(\Delta m^2_{31})$ and $m_0$.}

\begin{figure}[t]
\begin{center}
\includegraphics[origin=c, angle=-90, scale=0.3]{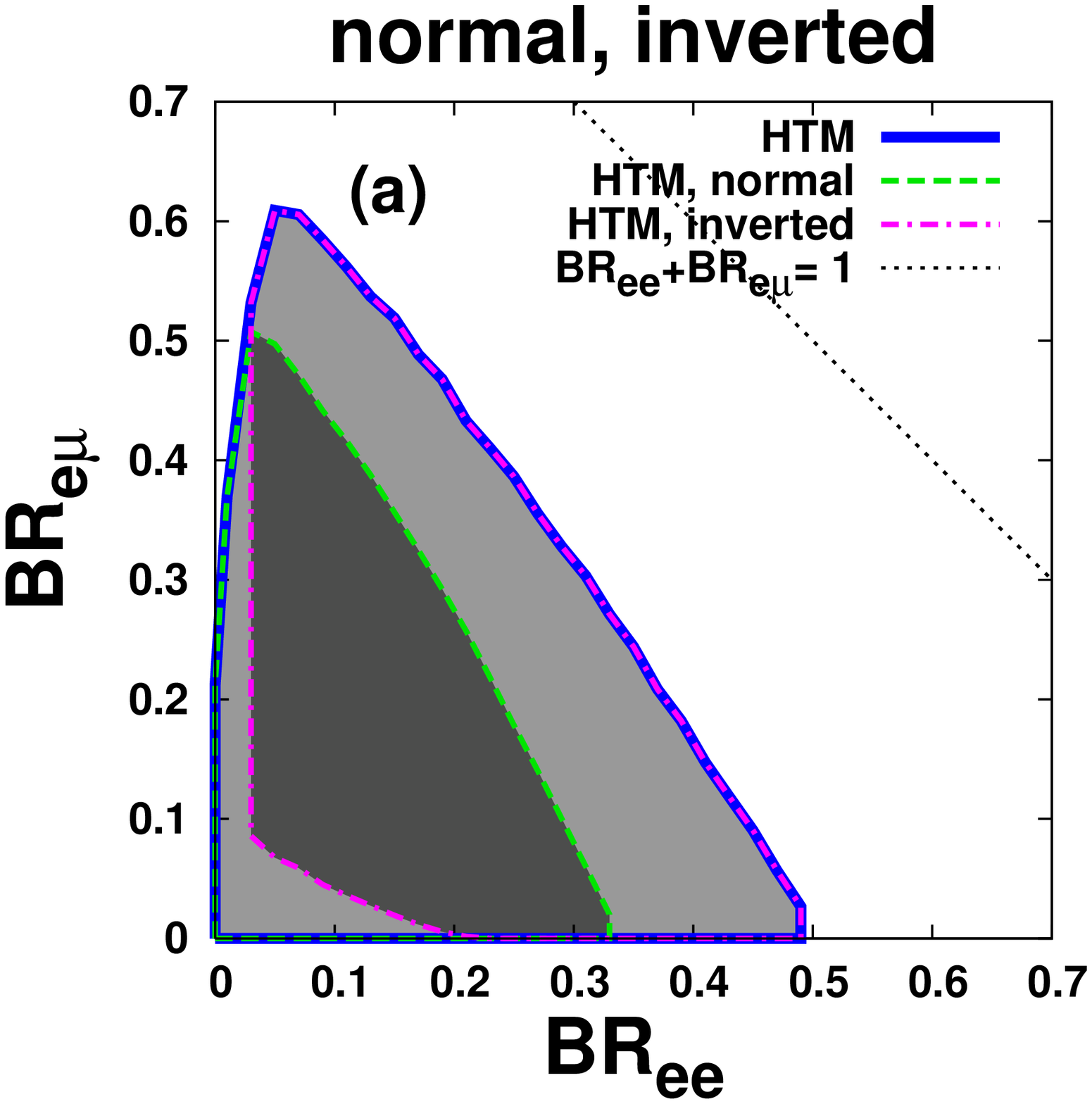}
\includegraphics[origin=c, angle=-90, scale=0.3]{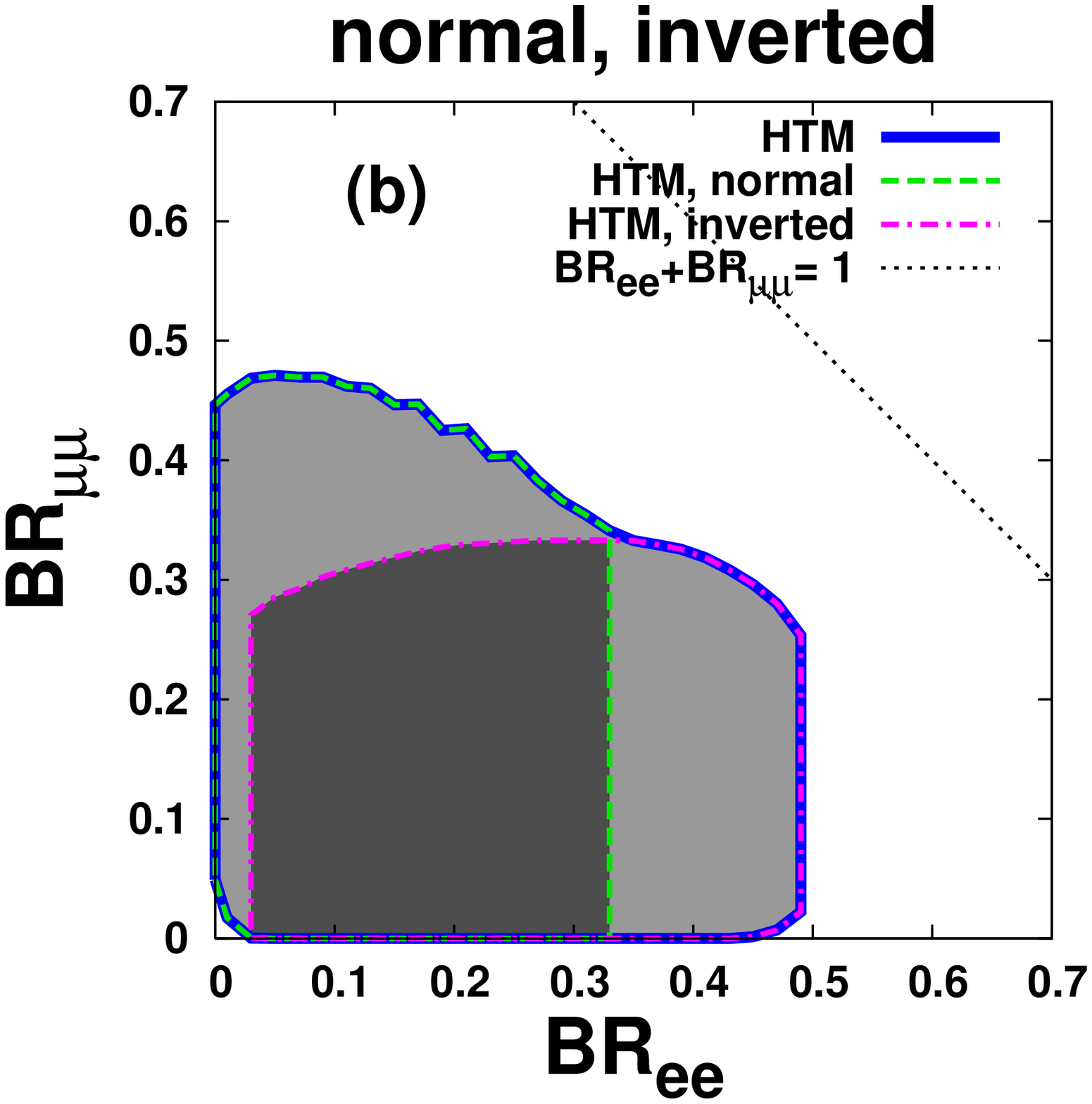}
\includegraphics[origin=c, angle=-90, scale=0.3]{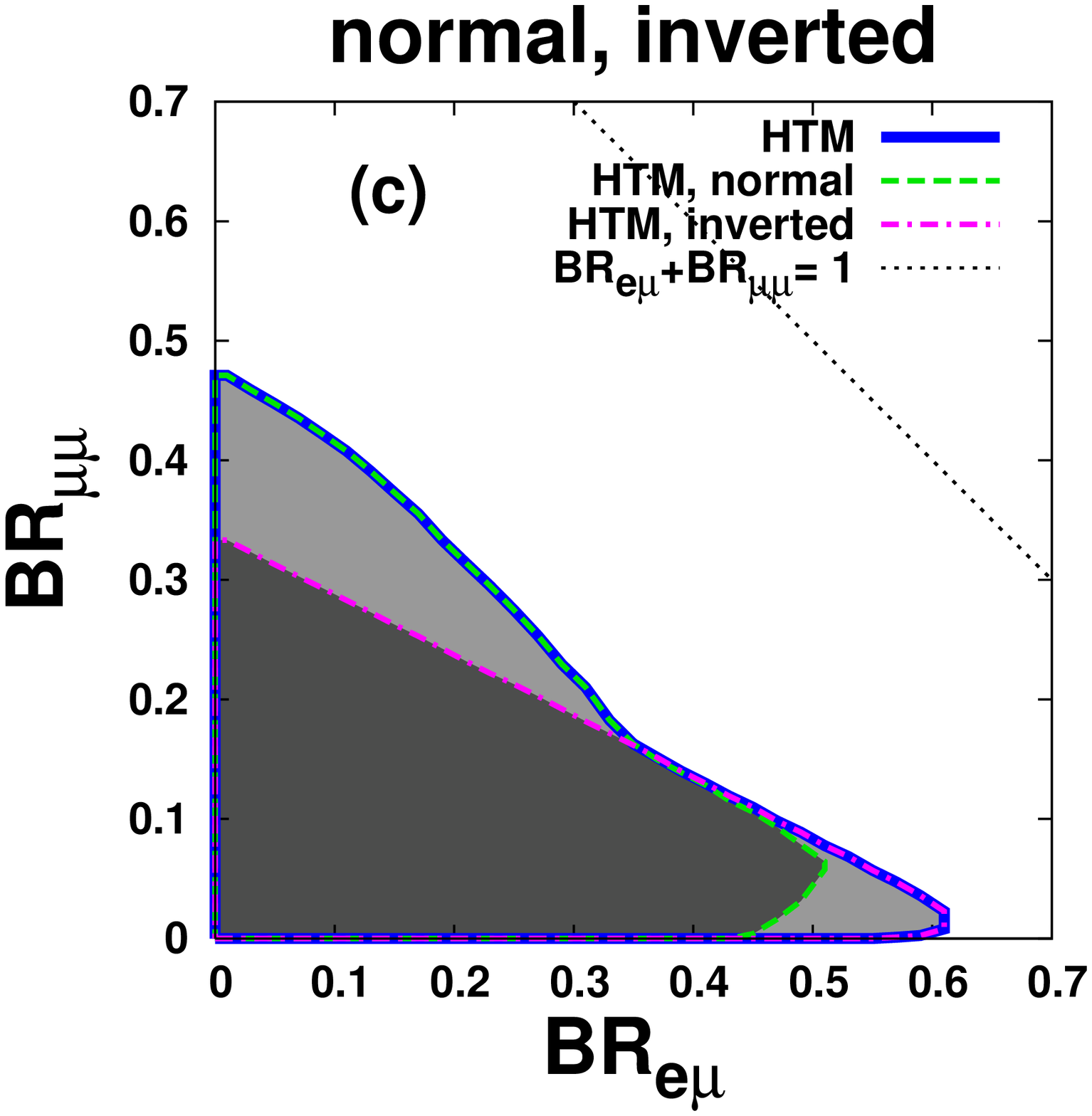}
\caption{
The shaded area surrounded by the solid line is attainable in the HTM\@.
Allowed regions for NH and IH correspond to $\BR$ inside of the
dashed and dash-dotted lines, respectively.
The dotted line divides the physical and unphysical regions.
}
\label{fig:BRni}
\end{center}
\end{figure}

 As we have seen, both
the neutrino mass spectrum and the Majorana phases have large effects on $\BRlilj$ in 
the HTM, but the model also gives a clear prediction for $\BR_{ll^\prime}$.
Therefore we can expect to obtain some information on
the neutrino mass spectrum and/or the Majorana phases
by observing $\BR_{ll^\prime}$.
In our analysis,
we use $\BR(H^{\pm\pm}\to e^\pm e^\pm, e^\pm \mu^\pm, \mu^\pm \mu^\pm)$
for which the LHC expects greatest sensitivity.
Na\"{\i}vely,
three measurements of $\BR_{ll^\prime}$ are sufficient to extract
information on the three parameters.

 In this section we consider the possibility
to determine $\text{sign}(\Delta m^2_{31})$
and/or to exclude $m_0 = 0$.
 Hereafter,
we use (\ref{fig1_para1}) and
\begin{eqnarray}
\sin^22\theta_{23}> 0.93 \,,~~~~
\sin^22\theta_{13}< 0.13\,,~~~~~ \delta=0\,\text{-}\,2\pi. 
\end{eqnarray}
 Non-zero values of $\theta_{13}$ and $\delta$ affect
$\BR_{ee}$ as was shown in Fig.~\ref{fig2}.
Moreover, deviation of $\theta_{23}$ from $\pi/4$
especially affects $\BR_{\mu\mu}$ in our analysis
because a rather wide range $0.37 < s_{23}^2 < 0.63$ is allowed. 
 In Fig.~\ref{fig:BRni} the
allowed regions of $\BR_{ll^\prime}$ in the HTM are shown
by the shaded (light and dark) regions; the dark shaded
area corresponds to the overlap of the allowed
regions for NH and IH\@. The area above the dotted line is unphysical
because the sum of $\BR$ exceeds unity.
It is clear that
the HTM predicts $\BR_{ee} \lesssim 0.49$,
$\BR_{e\mu} \lesssim 0.61$, and $\BR_{\mu\mu} \lesssim 0.47$.
The areas outside the solid line are unreachable,
and are particular to the HTM
(in which neutrino mass is solely given by eq.~(\ref{nu_mass}))
and will differ from the corresponding disallowed regions, 
if any,\footnote{If $h_{ij}$ are arbitrary parameters,
as in the Left-Right symmetric model of \cite{Mohapatra:1979ia},
the unreachable region would vanish.} 
in other models which can contain a light $H^{\pm\pm}$. 
 Hence in the event of signals for 
$\Hlilj$ these allowed regions can be compared
with the experimental measurements of $\BRlilj$
in order to test the HTM\@.
If measured $\BR$ are outside the shaded region
the HTM is disfavored.
If given signals for $\Hlilj$ 
are compatible with the reachable regions in the HTM,
the data can be interpreted in the context of this model and
one can try to extract information on parameters
of neutrino mass matrix.

Furthermore,
the area inside the dashed and dash-dotted lines in Fig.~\ref{fig:BRni}
is possible for NH and IH, respectively.
NH gives the additional bounds
$\BR_{ee} \lesssim 0.33$ and $\BR_{\mu\mu} \lesssim 0.51$, while
IH gives $0.03 \lesssim \BR_{ee}$ and $\BR_{\mu\mu} \lesssim 0.34$.
 These are also particular features of the HTM\@.
 If $\BR$ in the light shaded region are measured,
NH or IH is disfavored in the HTM\@.

\begin{figure}[t]
\begin{center}
\includegraphics[origin=c, angle=-90, scale=0.3]{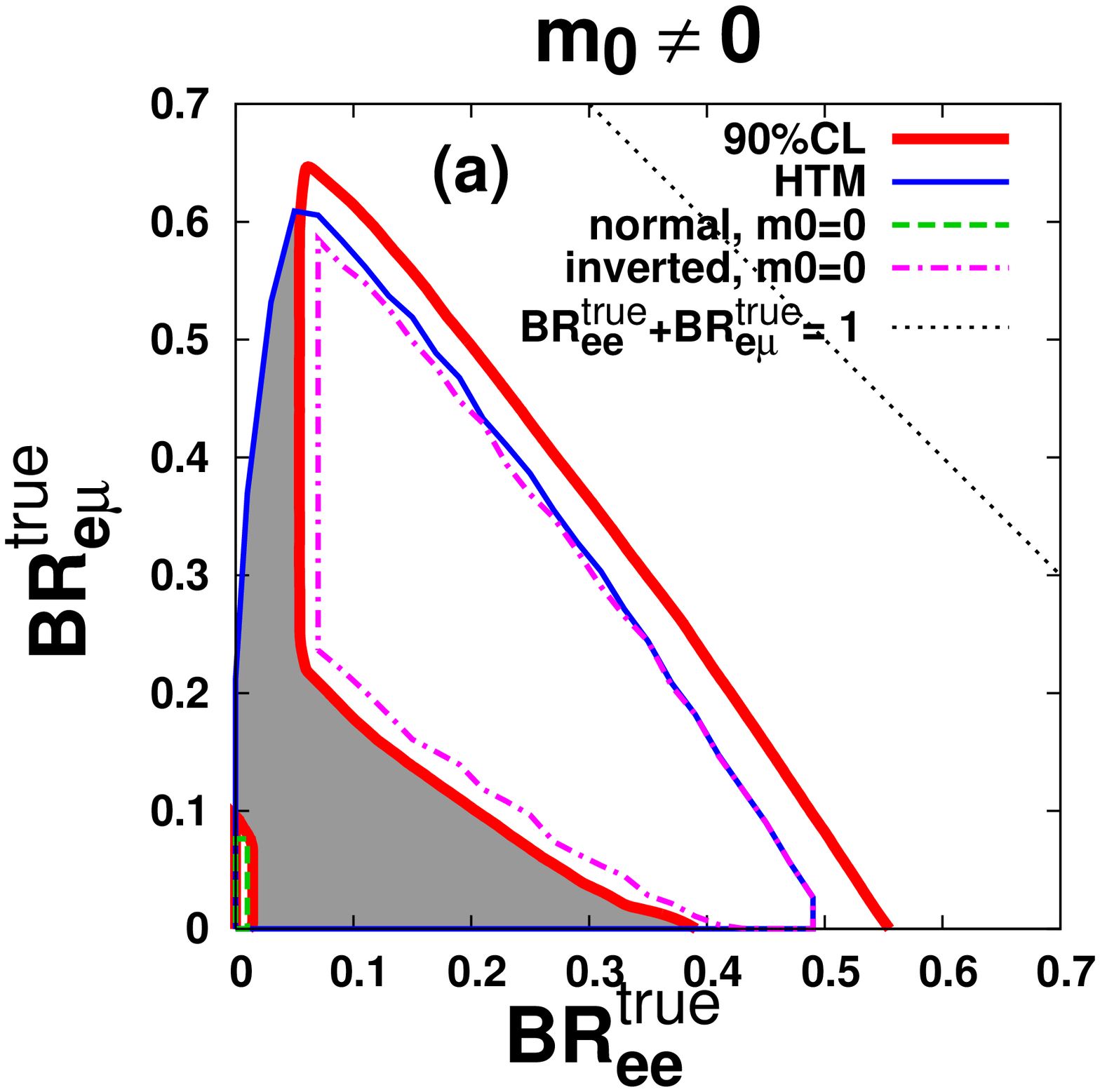}
\includegraphics[origin=c, angle=-90, scale=0.3]{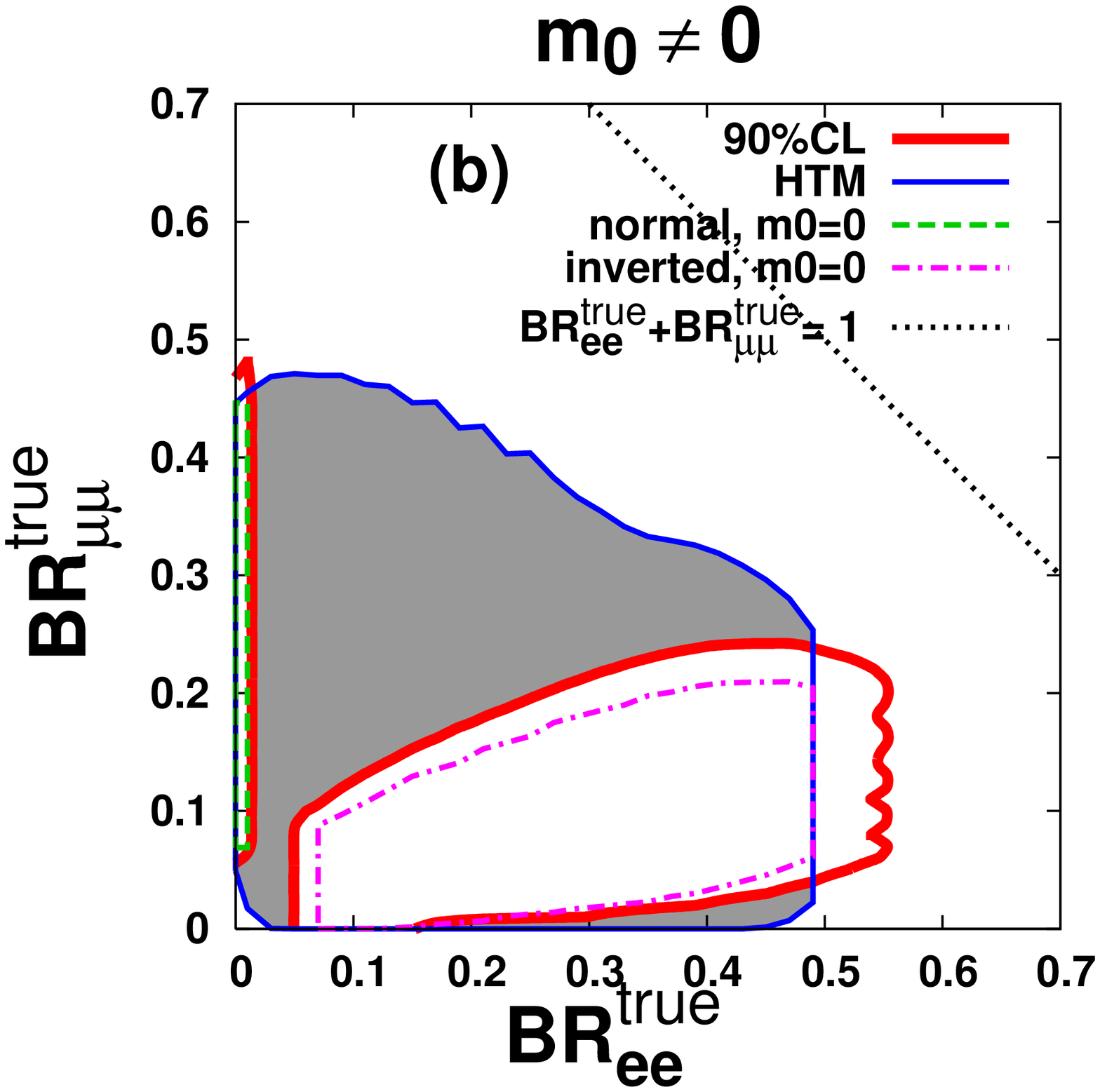}
\includegraphics[origin=c, angle=-90, scale=0.3]{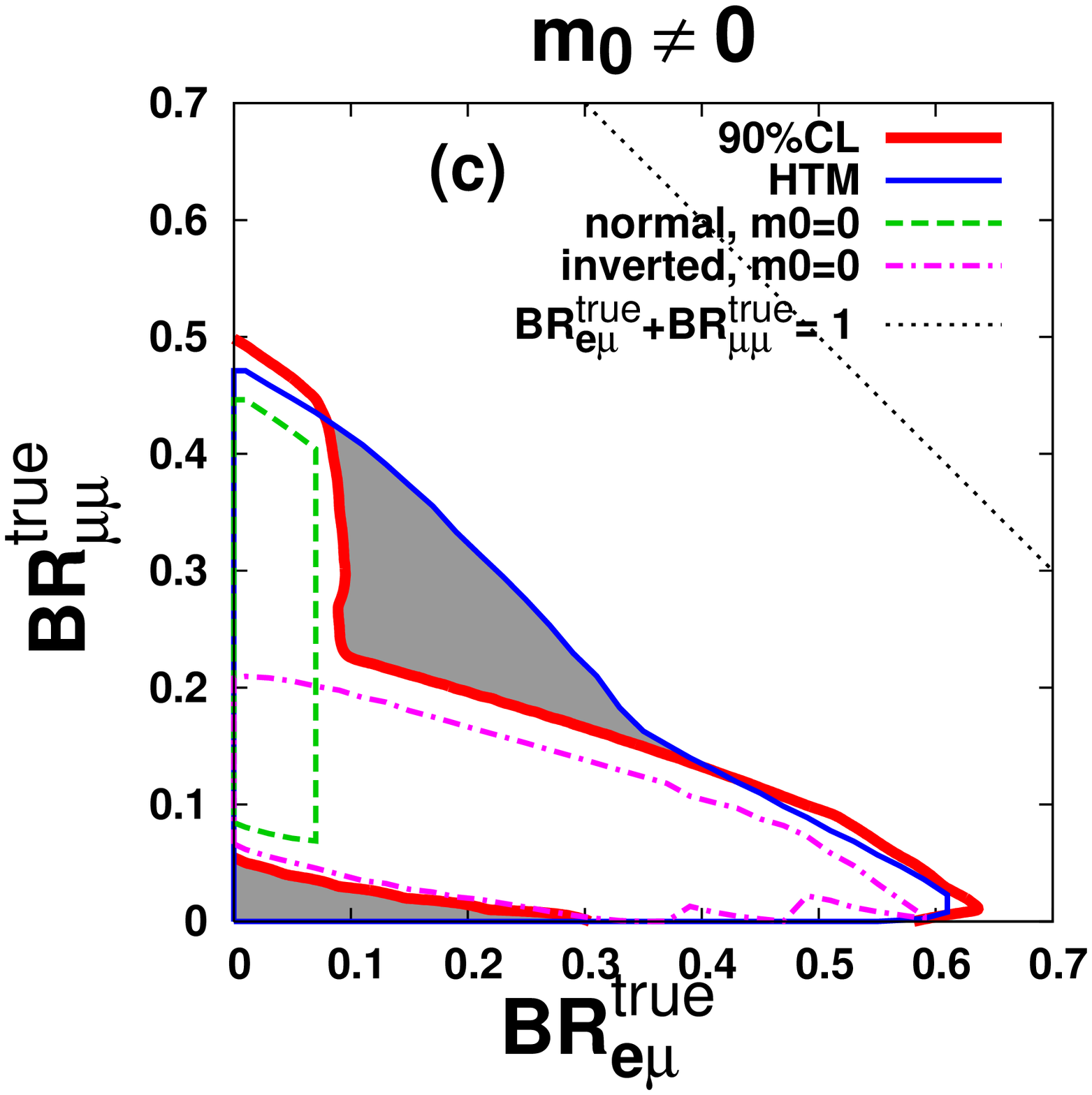}
\caption{
The thin solid line shows the attainable region in the HTM
and the dotted line divides the physical and unphysical regions.
The area inside the dashed and dash-dotted lines can be obtained
with $m_0=0$ in NH and IH, respectively.
The bold solid line is a result of a $\chi^2$ analysis
and $m_0=0$ can be excluded if $\BR$ are measured
within the shaded region.
}
\label{fig:BRmass}
\end{center}
\end{figure}

 Next, let us consider the possibility to exclude $m_0 = 0$.
 Fig.~\ref{fig:BRmass} shows the attainable regions 
of $\BR$ with $m_0 = 0$.
As in Fig.~\ref{fig:BRni}, the area inside the thin solid line is 
reachable in the HTM, and the region above the
dotted line is unphysical.
The dashed line corresponds to the $\BR$
which can be obtained in NH,
and the dash-dotted line is  $\BR$ in IH, both with $m_0 = 0$.
Note that the area inside these lines can also be obtained
with $m_0\neq 0$ but outside is impossible for $m_0=0$.
 This behaviour can also be seen in Fig.~\ref{fig3}.
The bold solid line denotes a $\chi^2$ analysis
for measurements of signals in the $2l$ channels
($N_{ee}$, $N_{e\mu}$, and $N_{\mu\mu}$)
of $H^{\pm\pm}$ decays
(see Appendix~\ref{sec:chi} for detail).
 If $\BR$ in the shaded region are measured,
one can exclude $m_0=0$ at 90\%CL in HTM\@.
 NH with $m_0=0$ gives a very clear prediction
of small $\BR_{ee}^\true$ as can be seen in Fig.~\ref{fig3}.
The bold line shows that
$m_0=0$ can be excluded at 90\%CL if $\BR_{ee}^\true$
lies in the narrow interval $0.02 \lesssim \BR_{ee}^\true \lesssim 0.05$,
even without measuring other $\BR$.
If only $\BR_{e\mu}$ or $\BR_{\mu\mu}$ is observed
it is impossible to exclude $m_0=0$. Therefore,
it is essential to use more than two $\BR$ to exclude $m_0=0$.

\subsection{Sensitivity to CP violation from Majorana phases}

\begin{figure}[t]
\begin{center}
\includegraphics[origin=c, angle=-90, scale=0.3]{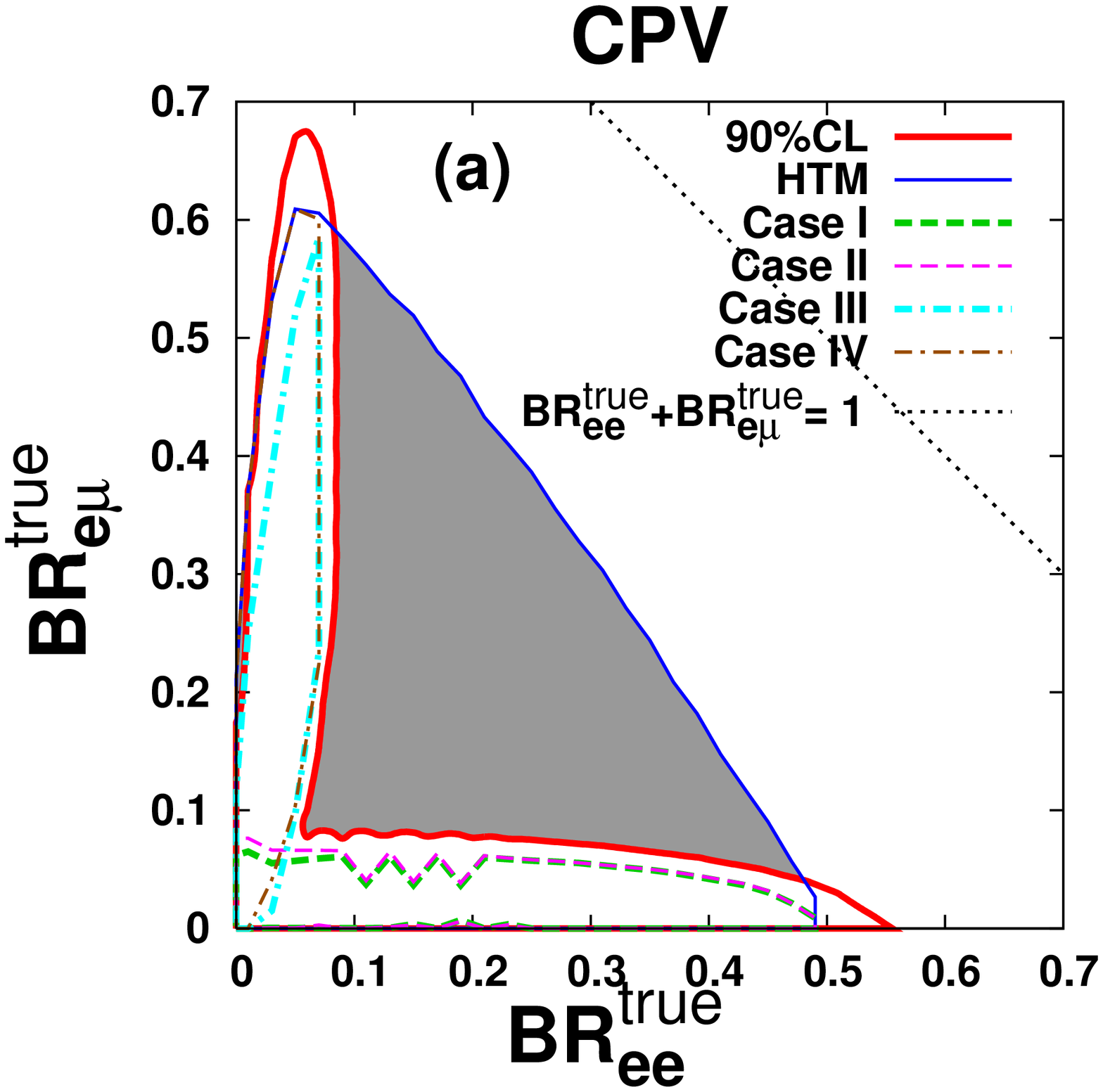}
\includegraphics[origin=c, angle=-90, scale=0.3]{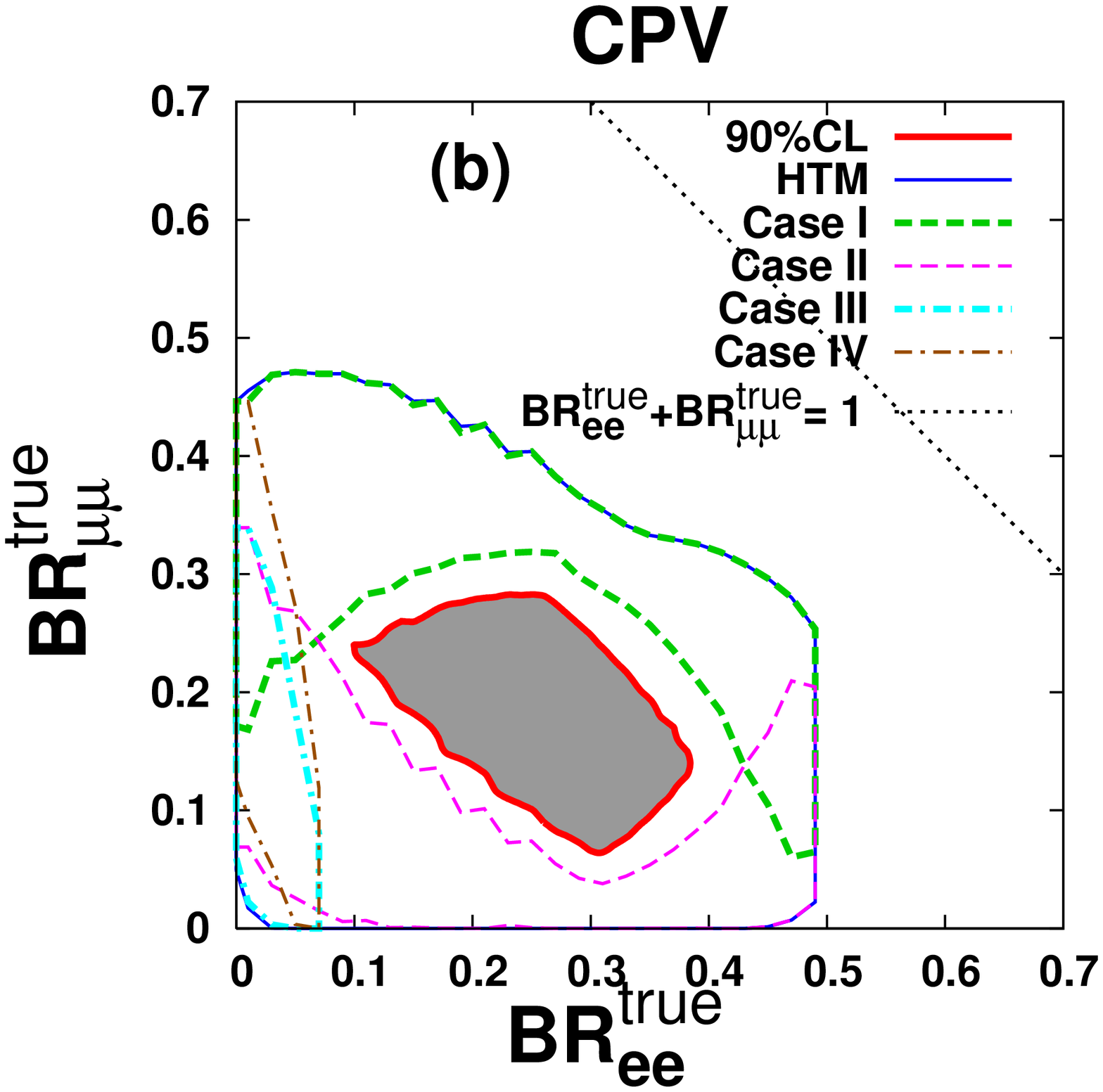}
\includegraphics[origin=c, angle=-90, scale=0.3]{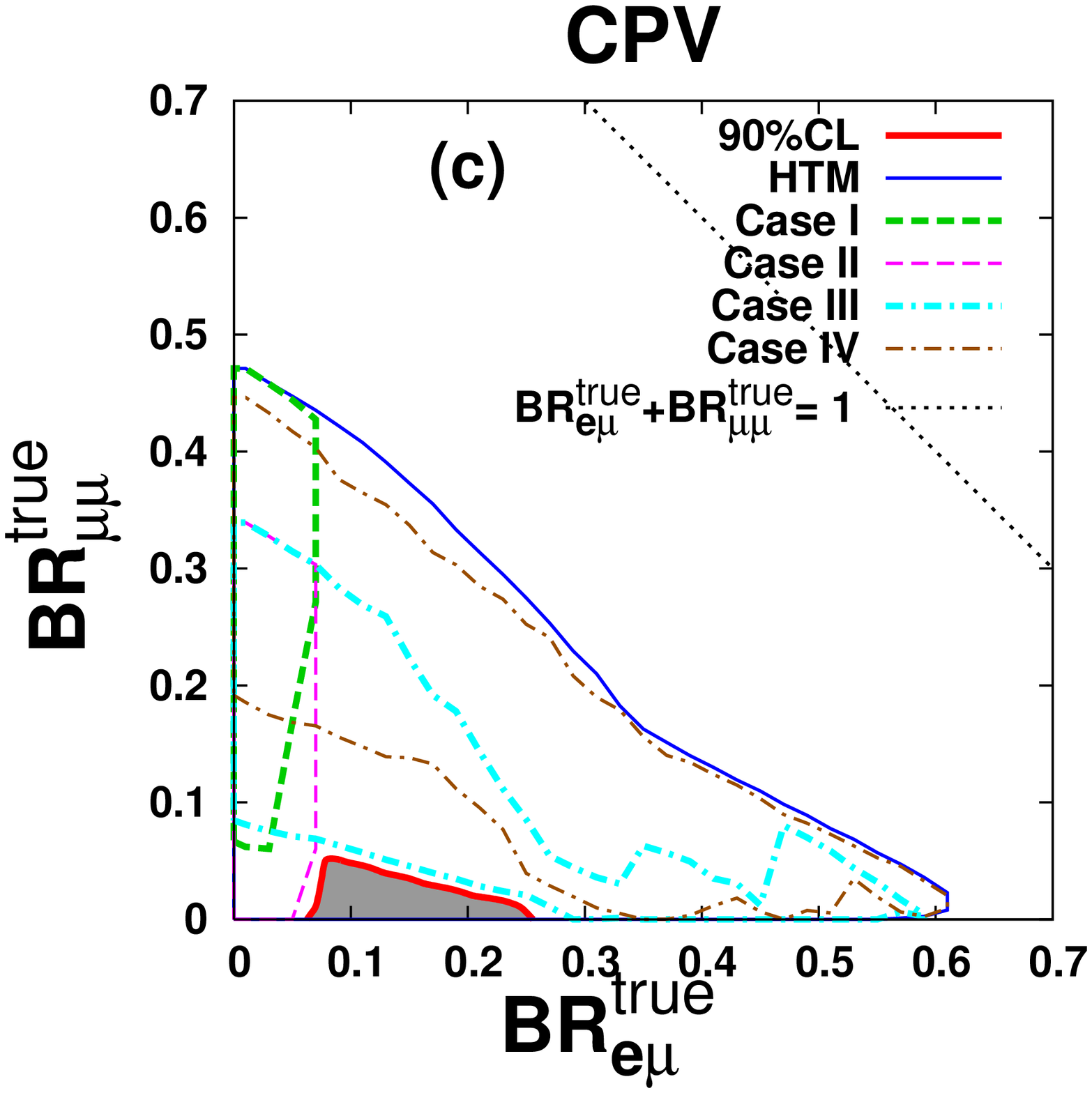}
\caption{
The thin solid line shows the attainable region in the HTM
and the dotted line divides the physical and unphysical regions.
Cases~I, II, III, and IV with CP conservation for Majorana phases
can give $\BR$ within the bold dashed, thin dashed,
bold dash-dotted, and thin dash-dotted lines, respectively.
The bold solid line is a result of a $\chi^2$ analysis
and the CP-conserving case is excluded at 90\%CL
if $\BR$ within the shaded region are measured.
}
\label{fig:BRcpv}
\end{center}
\end{figure}

In this section, we discuss the possibility to exclude
the CP-conserving cases for Majorana phases,
namely $\varphi_1, \varphi_2 = 0, \pi$.
 Establishing experimentally that
CP is violated by Majorana phases
in the HTM is of much phenomenological interest
and is feasible if there are sufficiently large numbers of
$H^{\pm\pm}$.

 In Fig.~\ref{fig:BRcpv},
the attainable regions of $\BR$ for the
CP-conserving cases of Majorana phases are shown.
As in Fig.~\ref{fig:BRni},
the area inside the thin solid line is reachable in the HTM,
and the region above the dotted line is unphysical.
Bold dashed, thin dashed, bold dash-dotted,
and thin dash-dotted lines show
the possible regions for Case~I, II, II, and IV, respectively.
From the figures it is clear that the HTM predicts very
specific regions for $\BRlilj$ in the CP-conserving cases,
especially in the space of $\BR_{ee}\text{-}\BR_{e\mu}$;
$\BR_{ee}$ is small ($\lesssim 0.07$)
in Case~III and IV (cases with $\varphi_1 = \pi$)
as seen in Fig.~\ref{fig2},
and $\BR_{e\mu}$ cannot be so large ($\lesssim 0.09$)
in Case~I and II (cases with $\varphi_1 = 0$).
The bold solid line is obtained by a $\chi^2$ analysis
with $2l$ channels of $H^{\pm\pm}$ decays
(see Appendix~\ref{sec:chi} for details),
and the CP-conserving cases are excluded at 90\%CL in the HTM
if experiment measures $\BR$ in the shaded region.

\subsection{Accuracy of the determination of $m_0$ and Majorana phases}

\begin{figure}[t]
\begin{center}
\includegraphics[origin=c, angle=-90, scale=0.29]{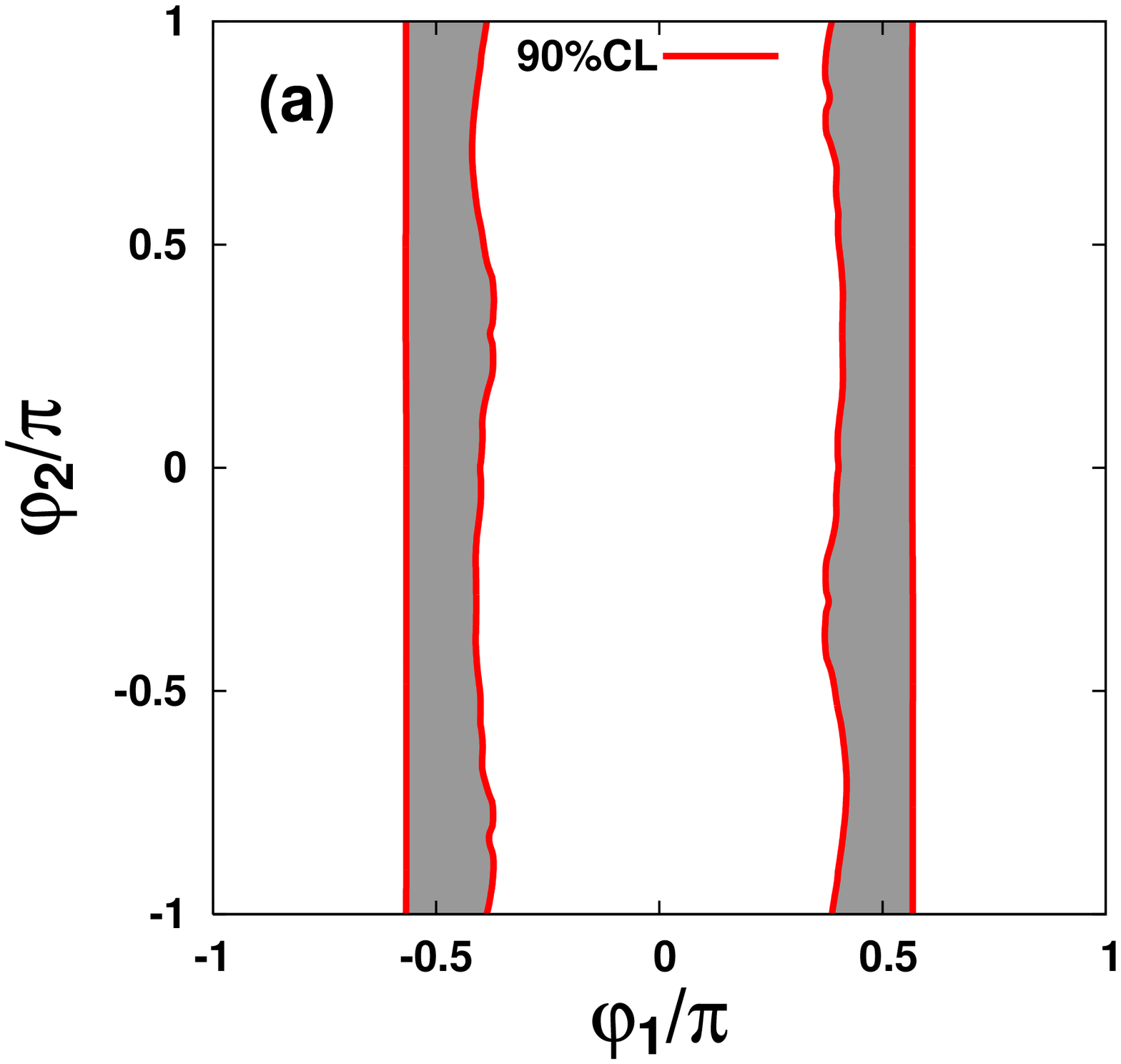}
\includegraphics[origin=c, angle=-90, scale=0.29]{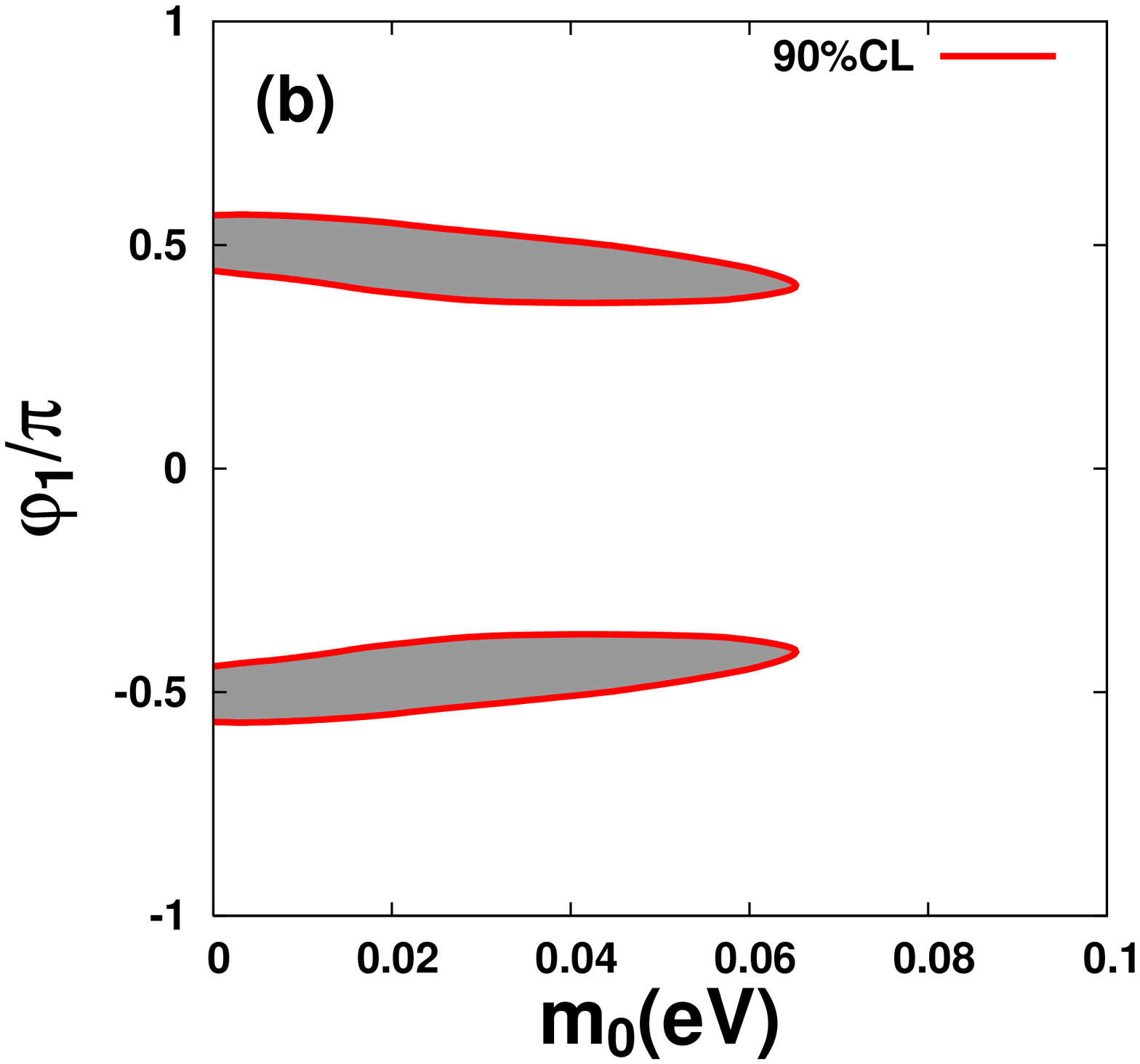}
\includegraphics[origin=c, angle=-90, scale=0.29]{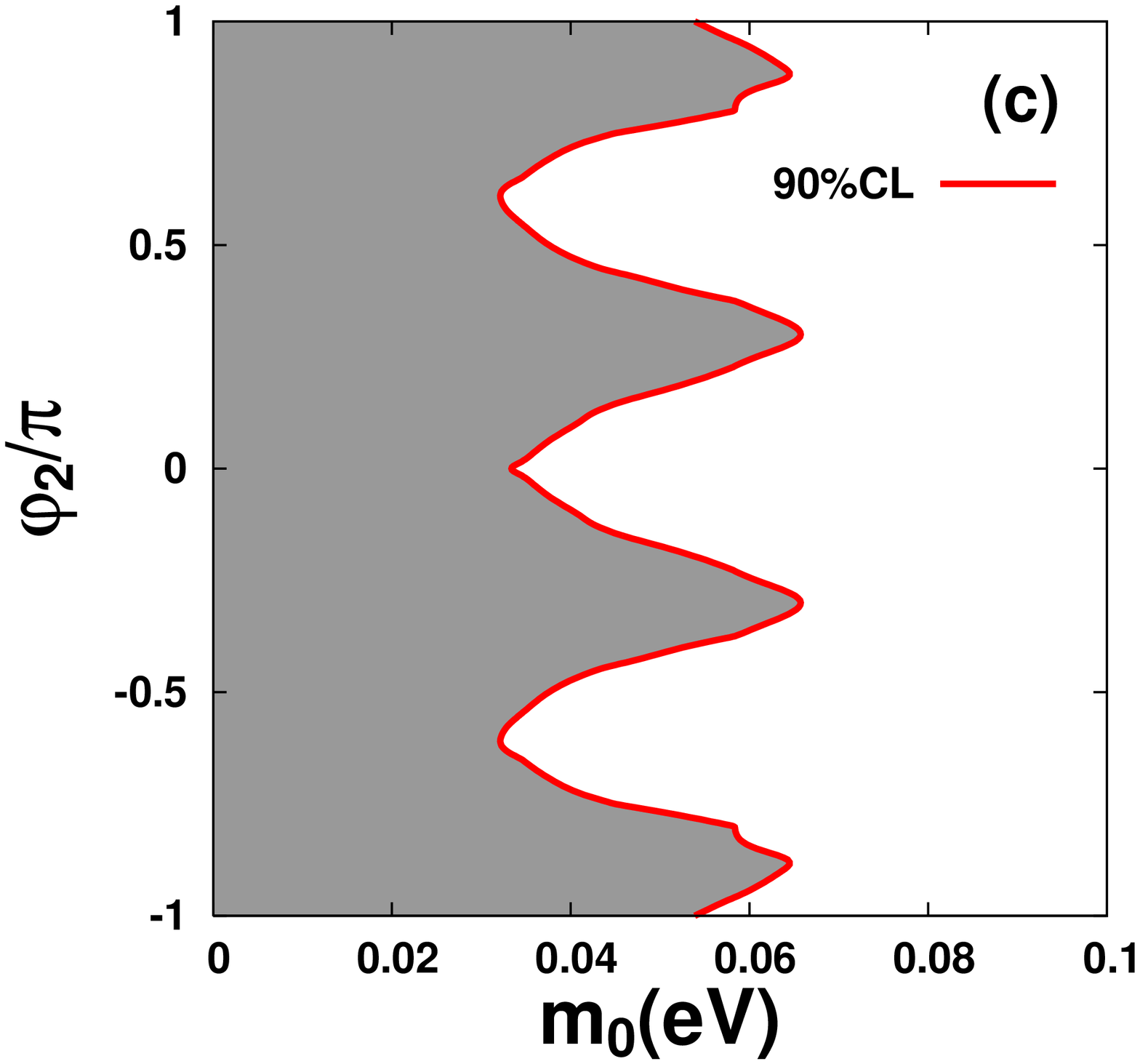}
\caption{
The shaded region is allowed at 90\%CL
for $\BR_{ee} = 0.268$, $\BR_{e\mu} = 0.243$,
and $\BR_{\mu\mu} = 0.066$.
These values can be obtained
by taking $m_0 = 0$, $\varphi_1 = \pi/2$,
$\sin^2{2\theta_{13}}=0.1$, $\delta=0$,
and $\theta_{23} = \pi/4$ for IH in HTM.
}
\label{fig:allowed}
\end{center}
\end{figure}

Once measurements of
$\BR_{ee}$, $\BR_{e\mu}$, and $\BR_{\mu\mu}$ are available
one can constrain $m_0$ and the Majorana phases.
 In this section
we show how precisely they can be constrained
by assuming $\BR_{ee} = 0.268$, $\BR_{e\mu} = 0.243$,
and $\BR_{\mu\mu} = 0.066$ as an example of a possible signal.
These values can be obtained
by taking $m_0 = 0$, $\varphi_1 = \pi/2$,
$\sin^2{2\theta_{13}}=0.1$, $\delta=0$,
and $\theta_{23} = \pi/4$ for IH in HTM\@.
It is evident from Fig.~\ref{fig:BRni} that
these values of $\BR$ can be accommodated in the HTM,
but not in the case of NH - see Fig.~\ref{fig:BRni}(a).

In Fig.~\ref{fig:allowed}, the
allowed regions at 90\%CL are shown
in the spaces of $\varphi_1$-$\varphi_2$, $m_0$-$\varphi_1$,
and $m_0$-$\varphi_2$.
They are obtained by a $\chi^2$ analysis
with three numbers of $2l$ channel signals
($N_{ee}$, $N_{e\mu}$, and $N_{\mu\mu}$)
of $H^{\pm\pm}$ decays
(See Appendix~\ref{sec:chi} for details).
Since the above $\BR$ are inside the shaded regions in Fig.~\ref{fig:BRcpv},
Case~I-IV of CP-conservation are all excluded which is clearly
depicted in Fig.~\ref{fig:allowed}(a).
 Two regions in Fig.~\ref{fig:allowed}(a) are just copys
of each other
because $\BR_{ll^\prime}$ are unchanged by replacing
$h_{ij}$ with $h_{ij}^\ast$.
Although $\varphi_1$ can be constrained very well,
all values of $\varphi_2$ are allowed in Fig.~\ref{fig:allowed}
because $\BR$ in IH with $m_0=0$ does not depend on $\varphi_2$.
In general one expects looser constraints on $\varphi_2$ 
because dependence of $\BR_{ee}$ and $\BR_{e\mu}$ on $\varphi_2$ 
is suppressed by tiny $\theta_{13}$.
Fig.~\ref{fig:allowed} also shows that
a stringent upper bound on $m_0$
($m_0 \lesssim 0.07$ in this example)
can be obtained by measuring $\BR$.

\section{Conclusions}
Doubly charged Higgs bosons ($H^{\pm\pm}$) are
a distinctive signature of a variety of
models which can accommodate neutrino mass. 
Discovery of $H^{\pm\pm}$ at the Tevatron/LHC with a reasonably
light mass ($< 400$~GeV) could enable thousands of  
$\Hlilj$ events to be observed with the anticipated final integrated 
luminosities at the LHC\@.
In such a scenario precise measurements of 
$\Hlilj$ would be feasible and 
would constitute a means of experimentally favouring or
disfavouring various candidate models which contain a $H^{\pm\pm}$.

The Higgs Triplet Model (HTM), in which (Majorana) neutrino mass
is solely given as the product of a triplet vacuum expectation value
$v_{\Delta}$ and a triplet Yukawa coupling $h_{ij}$,
is a particularly simple model which leads to a specific relationship
between the branching ratios (BR) of $\Hlilj$
and the neutrino mass matrix.
In particular,
BRs are determined only by the neutrino mass matrix
independently of $v_\Delta$ and Higgs boson masses
if $m_{H^{\pm\pm}} \leq m_{H^\pm}$.
We performed a quantitative study of the dependence of the BRs on
all the parameters in the neutrino mass matrix,
especially the dependence on the 
three parameters which induce the most uncertainty
in such predictions: 
the lightest neutrino mass $m_0$ and the two Majorana phases.
 We displayed the allowed regions of $\BRlilj$
for $i, j = e,\mu$ and showed that they are considerably smaller
than for the case of an arbitrary $h_{ij}$.
By measuring these BRs it is possible to both test the HTM and
determine the sign of $\Delta m^2_{31}$.

Due to the large effect of $m_0$ and the Majorana phases on 
the BRs we showed that it is possible to extract information
on these three parameters by measuring
$\BR_{ee}$, $\BR_{e\mu}$, and $\BR_{\mu\mu}$.
Such information cannot be obtained from
neutrino oscillation experiments.
By using the $2l$ channel of $H^{\pm\pm}$ decays,
it was found that $m_0=0$ and/or the 
CP-conserving case for the Majorana phases can be excluded
without knowledge of $\text{sign}(\Delta m^2_{31})$,
$m_0$, $\theta_{13}$, and $\delta$.
There is a small range of $\BR_{ee}$
($0.02 \lesssim \BR_{ee} \lesssim 0.05$)
where $m_0=0$ can be excluded at 90\%CL
by measuring only $\BR_{ee}$.
For the exclusion of the CP-conserving case for Majorana phases,
$\BR_{ee}$ and $\BR_{e\mu}$ are especially important.
For a specific choice of $\BR_{ee}$, $\BR_{e\mu}$, and $\BR_{\mu\mu}$
we displayed the allowed regions
of $(m_0, \varphi_1, \varphi_2)$ and showed
that it is possible to constrain $m_0$ and $\varphi_1$ stringently.

In contrast, establishing that Majorana phases are non-zero
is notoriously difficult in models in which lepton number violation is 
associated with particles at a very high energy scale
(e.g. right-handed neutrinos with a Majorana mass of order
$10^{10}~\GeV$ or more). In the HTM  with a light $H^{\pm\pm}$
this becomes is a realistic and unique possibility.

 Note added:
During finalization of this paper
an article \cite{Garayoa:2007fw} appeared which deals
with the same topic. Although there is inevitably some overlap, our
analysis differs considerably from theirs
e.g., i) our definition of the $\chi^2$ is a function of three BRs in the
2 lepton channel, while their $\chi^2$ is for five BRs in the 4 lepton
channel, and ii) the presentation of results, especially figures 5-8.
Where comparisons can be made we find good agreement with their results.

\section*{Acknowledgements}
A.G.A. was supported by National Cheng Kung University Grant 
No.\ OUA 95-3-2-057.
 The work of M.A. was supported, in part,
by Japan Society for the Promotion of Science.

\appendix
\section{Definition of $\Delta \chi^2$}\label{sec:chi}

In order to test assumptions,
e.g., CP-conservation of Majorana phases, we calculate a $\Delta\chi^2$.
 In our analysis,
$\BR_{ee}^\true$, $\BR_{e\mu}^\true$, and $\BR_{\mu\mu}^\true$
are used as input which can be understood as
true values chosen by nature or best fit values from experiment.
 We try to fit these true values with $\BR_{ll^\prime}^\fit$
which are calculated theoretically by eq.~(\ref{BRll})
for the assumption to be tested.
 We utilize the $2l$ channel
and the number of events is given by
\begin{eqnarray}
N_{ll^\prime} \equiv
\epsilon_\eff \left\{
               N_{\text{pair}} \BR_{ll^\prime} (2-\BR_{ll^\prime})
               + N_{\text{single}} \BR_{ll^\prime}
              \right\},
\end{eqnarray}
where $N_{\text{pair}}$ ($N_{\text{single}}$)
is the number of pair (single) production of doubly charged Higgs,
and $\epsilon_\eff$ denotes an efficiency due to event cuts;
 we can obtain the numbers of $H^{\pm\pm}$ production
from Table~\ref{event_no} as
$N_{\text{pair}} = N_{4l}$ and
$N_{\text{single}} = N_{2l} - N_{4l}$.
 We take $\epsilon_\eff = 0.5$.
 We use $N_{\text{pair}} = 900$ and $N_{\text{single}} = 1600$
by assuming $m_{H^{\pm\pm}} = m_{H^\pm}$ for simplicity
(see Table~\ref{event_no}) and
because the difference of their masses should not be too large
in order to maintain $\rho \simeq 1$ at the 1-loop level.
For the number of remaining background events after cuts,
we use $N_\BG = 1$ for each $N_{ll^\prime}$
because the signal is expected to be almost background free.
 In the fitting procedure,
$\Delta \chi^2$ is minimized within possible regions
of the following parameters:
\begin{eqnarray}
m_0^\fit &=& 0\,\text{-}\,0.3 \eV\,, \
\varphi_1^\fit, \varphi_2^\fit = 0\,\text{-}\,2\pi,\\
\sin^22\theta_{13}^\fit &=& 0\,\text{-}\,0.13\,, \
\delta^\fit = 0\,\text{-}\,2\pi\,,\\
\sin^22\theta_{23}^\fit &>& 0.93\,, \ 
\hierarchy^\fit ={\rm NH}, {\rm IH}.
\label{fit_para}
\end{eqnarray}
 The minimization gives the most pessimistic value
for exclusion of the assumption
because a smaller value of $\Delta\chi^2$
means a better fitting with the assumption.
 The other parameters are fixed
for true and fitting values as (\ref{fig1_para1}).
Since we wish to extract information on three parameters
$(m_0, \varphi_1, \varphi_2)$,
our analysis is for three degrees of freedom
and $\Delta \chi^2 = 6.3$ corresponds to 90\%CL
even if results are projected onto spaces of two parameters.

\subsection{Exclusion of $m_0=0$ and CP-conserving case for Majorana phases}

 In Fig.~\ref{fig:BRmass},
the region where $m_0 = 0$ is excluded in the HTM
is given by calculating the $\Delta \chi^2_{m_0}$ function.
 The $\Delta \chi^2_{m_0}$ is defined as
\begin{eqnarray}
\Delta\chi^2_{m_0} ( \BR_{ee}^\true, \BR_{e\mu}^\true, \BR_{\mu\mu}^\true )&=&
 \min_{ \varphi_1^\fit, \varphi_2^\fit, x^\fit}
 \left\{\left.
  \sum_{ll^\prime = ee, e\mu, \mu\mu}
   \frac{
    \left(
     N_{ll^\prime}^\true - N_{ll^\prime}^\fit
    \right)^2
   }{
    N_{ll^\prime}^\true + N_\BG
    }
 \right|_{m_0^\fit =0}
 \right\},\\
\label{chi2m0}
x &\equiv&
\left\{ \theta_{13}, \delta, \theta_{23}, \hierarchy \right\}.
\end{eqnarray}
where $N_{ll^\prime}^\true$ is given by
$(\BR_{ee}^\true, \BR_{e\mu}^\true, \BR_{\mu\mu}^\true)$
and $N_{ll^\prime}^\fit$ is calculated with
the theoretical form of $\BR$~(eq.~(\ref{BRll}))
for $(\varphi_1^\fit, \varphi_2^\fit, m_0^\fit, x^\fit)$.
 When we show the projected region of
$\Delta \chi^2_{m_0} \leq 6.3$
onto the plane $\BR_{ee}^\true\text{-}\BR_{e\mu}^\true$,
for example,
we minimize the $\Delta\chi^2_{m_0}$
with respect to $\BR_{\mu\mu}^\true$;
 this minimization is achieved simply
by replacing $\mu\mu$ from the sum of $ll^\prime$
in the definition of the $\Delta \chi^2_{m_0}$,
and then we take 6.3
for the resulting $\Delta \chi^2 (\BR_{ee}^\fit, \BR_{e\mu}^\fit)$
to have the projected contour of 90\%CL\@.

 In Fig.~\ref{fig:BRcpv},
the region where all of the CP-conserving cases~I, II, III, and IV are
rejected is given by calculating the $\Delta \chi^2_\CPV$ function.
The $\Delta \chi^2_\CPV$ is
\begin{eqnarray}
\Delta\chi^2_\CPV ( \BR_{ee}^\true, \BR_{e\mu}^\true, \BR_{\mu\mu}^\true )&=&
 \min_{ \CPC, m_0^\fit, x^\fit}
 \left\{
  \sum_{ll^\prime =ee, e\mu, \mu\mu}
   \frac{
    \left(
     N_{ll^\prime}^\true - N_{ll^\prime}^\fit
    \right)^2
   }{
    N_{ll^\prime}^\true + N_\BG
    }
 \right\}.
\label{chi2cpv}
\end{eqnarray}
 We use $\Delta \chi^2_\CPV$ in the same way
as $\Delta \chi^2_{m_0}$.

\subsection{Allowed region of $\varphi_1$, $\varphi_2$, and $m_0$}

 In Fig.~\ref{fig:allowed},
the allowed region of $(\varphi_1, \varphi_2, m_0)$
for a given set of $\BR_{ll^\prime}^\true$ is obtained by 
\begin{eqnarray}
\Delta \chi^2_\allow (\varphi_1, \varphi_2, m_0)&=&
 \min_{x^\fit}
 \left\{
 \sum_{ll^\prime = ee, e\mu, \mu\mu}
  \frac{
   \left(
    N_{ll^\prime}^\true - N_{ll^\prime}^\fit
   \right)^2
  }{
    N_{ll^\prime}^\true + N_{\BG}
   }
 \right\}.
\end{eqnarray}
 $N_{ll^\prime}^\fit$ is a function of
$(\varphi_1, \varphi_2, m_0, x^\fit)$.
 In order to project the 3-dimensional allowed region
onto 2-dimensional space,
we minimize $\Delta \chi^2_\allow$ with respect to
a parameter.
 Then, we take 6.3 for the resulting $\Delta \chi^2$
to obtain the 90\%CL contour.

\end{document}